\documentstyle[12pt]{article}

\skewchar\fivmi='177
\skewchar\sixmi='177
\skewchar\sevmi='177
\skewchar\egtmi='177
\skewchar\ninmi='177
\skewchar\tenmi='177
\skewchar\elvmi='177
\skewchar\twlmi='177
\skewchar\frtnmi='177
\skewchar\svtnmi='177
\skewchar\twtymi='177
\def\@magscale#1{ scaled \magstep #1}
\skewchar\fivsy='60
\skewchar\sixsy='60
\skewchar\sevsy='60
\skewchar\egtsy='60
\skewchar\ninsy='60
\skewchar\tensy='60
\skewchar\elvsy='60
\skewchar\twlsy='60
\skewchar\frtnsy='60
\skewchar\svtnsy='60
\skewchar\twtysy='60

\catcode`@=11  
\def\un#1{\relax\ifmmode\@@underline#1\else
        $\@@underline{\hbox{#1}}$\relax\fi}
\catcode`@=12

\def\a{\alpha}
\def\b{\beta}

\def\d{\delta}
\def\e{\epsilon}

\def\g{\gamma}

\def\k{\kappa}
\def\l{\lambda}
\def\m{\mu}

\def\q{\theta}
\def\r{\rho}
\def\s{\sigma}

\def\z{\zeta}
\def\D{\Delta}

\def\G{\Gamma}

\def\L{\Lambda}
\def\O{\Omega}

\def\S{\Sigma}

\def\dslash{\not{\hbox{\kern-2pt $\partial$}}}
\def\Dslash{\not{\hbox{\kern-4pt $D$}}}
\def\pslash{\not{\hbox{\kern-2.3pt $p$}}}
 \newtoks\slashfraction
 \slashfraction={.13}
 \def\slash#1{\setbox0\hbox{$ #1 $}
 \setbox0\hbox to \the\slashfraction\wd0{\hss \box0}/\box0 }
 
\font\ro=cmsy10                          
\def\kcr{{\hbox{\ro \char'170}}}                
\def\ktl{{\hbox{\ro \char'170}}}        
\def\ktr{{\hbox{\ro \char'170}}}        
\def\kbl{{\hbox{\ro \char'170}}}        
\def\kbr{{\hbox{\ro \char'170}}}        

\def\plpl{\raise-2pt\hbox{$\raise3pt\hbox{$_+$}\hskip-6.67pt\raise0.0pt
\hbox{$^+$}\hskip 0.01pt$}}
\def\mimi{\raise-2pt\hbox{$\raise3pt\hbox{$_-$}\hskip-6.67pt\raise0.0pt
\hbox{$^-$}\hskip 0.01pt$}} 

\def\bo{{\raise.15ex\hbox{\large$\Box$}}}               
\def\pa{\partial}                                       
\def\TH{{\raise.2ex\hbox{$\displaystyle \bigodot$}\mskip-4.7mu \llap H \;}}
\def\face{{\raise.2ex\hbox{$\displaystyle \bigodot$}\mskip-2.2mu \llap {$\ddot
        \smile$}}}                                      

\def\sp#1{{}^{#1}}                              
\def\Tilde#1{\widetilde{#1}}                    
\def\Hat#1{\widehat{#1}}                        
\def\Bar#1{\overline{#1}}                       
\def\leftrightarrowfill{$\mathsurround=0pt \mathord\leftarrow \mkern-6mu
        \cleaders\hbox{$\mkern-2mu \mathord- \mkern-2mu$}\hfill
        \mkern-6mu \mathord\rightarrow$}
\def\dvec#1{\vbox{\ialign{##\crcr
        \leftrightarrowfill\crcr\noalign{\kern-1pt\nointerlineskip}
        $\hfil\displaystyle{#1}\hfil$\crcr}}}           
\def\der#1{{\pa \over \pa {#1}}}                

\def\frac#1#2{{\textstyle{#1\over\vphantom2\smash{\raise.20ex
        \hbox{$\scriptstyle{#2}$}}}}}                   
\def\ha{\frac12}                                        
\def\sfrac#1#2{{\vphantom1\smash{\lower.5ex\hbox{\small$#1$}}\over
        \vphantom1\smash{\raise.4ex\hbox{\small$#2$}}}} 
\def\bfrac#1#2{{\vphantom1\smash{\lower.5ex\hbox{$#1$}}\over
        \vphantom1\smash{\raise.3ex\hbox{$#2$}}}}       
\def\afrac#1#2{{\vphantom1\smash{\lower.5ex\hbox{$#1$}}\over#2}}    
\def\partder#1#2{{\partial #1\over\partial #2}}   

\newskip\humongous \humongous=0pt plus 1000pt minus 1000pt
\def\caja{\mathsurround=0pt}
\def\eqalign#1{\,\vcenter{\openup2\jot \caja
        \ialign{\strut \hfil$\displaystyle{##}$&$
        \displaystyle{{}##}$\hfil\crcr#1\crcr}}\,}
\newif\ifdtup

\def\ref#1{$\sp{#1)}$}

\parskip=\medskipamount                 
\thicklines                         

\def\oldheadpic{                                
        \setlength{\unitlength}{.4mm}
        \thinlines
        \par
        \begin{picture}(349,16)
        \put(325,16){\line(1,0){4}}
        \put(330,16){\line(1,0){4}}
        \put(340,16){\line(1,0){4}}
        \put(335,0){\line(1,0){4}}
        \put(340,0){\line(1,0){4}}
        \put(345,0){\line(1,0){4}}
        \put(329,0){\line(0,1){16}}
        \put(330,0){\line(0,1){16}}
        \put(339,0){\line(0,1){16}}
        \put(340,0){\line(0,1){16}}
        \put(344,0){\line(0,1){16}}
        \put(345,0){\line(0,1){16}}
        \put(329,16){\oval(8,32)[bl]}
        \put(330,16){\oval(8,32)[br]}
        \put(339,0){\oval(8,32)[tl]}
        \put(345,0){\oval(8,32)[tr]}
        \end{picture}
        \par
        \thicklines
        \vskip.2in}
\def\oldtitle#1#2#3#4{\oldheadpic\begin{center}\vglue.5in{\large\bf #1}\\[.6in]
        {#2}\\[.1in] {\it Department of Physics and Astronomy}\\
        {\it University of Maryland, College Park, MD 20742}\\[.6in]
        Physics Publication \#{#3}\\ {#4}\\[1.5in] {\bf ABSTRACT}\\[.1in]
        \end{center} \begin{quotation}}                 
\def\oldTitle#1#2#3#4#5#6#7{\oldheadpic\begin{center} \vglue .4in
        {\large\bf #1}\\[.4in]
        {#2}\\[.1in] {\it Department of Physics and Astronomy}\\
        {\it University of Maryland, College Park, MD 20742}\\[.1in]
        {#3}\\[.1in] {\it {#4}}\\ {\it {#5}}\\[.4in]
        Physics Publication \#{#6}\\ {#7}\\[.5in] {\bf ABSTRACT}\\[.1in]
        \end{center} \begin{quotation}}                 
\def\border{                                            
        \setlength{\unitlength}{1mm}
        \newcount\xco
        \newcount\yco
        \xco=-21
        \yco=12
        \begin{picture}(140,0)
        \put(\xco,\yco){$\ktl$}
        \advance\yco by-1
        {\loop
        \put(\xco,\yco){$\kcr$}
        \advance\yco by-2
        \ifnum\yco>-240
        \repeat
        \put(\xco,\yco){$\kbl$}}
        \xco=158
        \yco=12
        \put(\xco,\yco){$\ktr$}
        \advance\yco by-1
        {\loop
        \put(\xco,\yco){$\kcr$}
        \advance\yco by-2
        \ifnum\yco>-240
        \repeat
        \put(\xco,\yco){$\kbr$}}
        \put(-20,13){\tiny University of Maryland Elementary Particle
Physics University of Maryland Elementary Particle Physics University of
Maryland Elementary Particle Physics}
        \put(-20,-241.5){\tiny University of Maryland Elementary
Particle Physics University of Maryland Elementary Particle Physics
University of Maryland Elementary Particle Physics}
        \end{picture}
        \par\vskip-8mm}
\def\bordero{                                           
        \setlength{\unitlength}{1mm}
        \newcount\xco
        \newcount\yco
        \xco=-31
        \yco=12
        \begin{picture}(140,0)
        \put(\xco,\yco){$\ktl$}
        \advance\yco by-1
        {\loop
        \put(\xco,\yco){$\kclr}
        \advance\yco by-2
        \ifnum\yco>-240
        \repeat
        \put(\xco,\yco){$\kbl$}}
        \xco=151
        \yco=12
        \put(\xco,\yco){$\ktr$}
        \advance\yco by-1
        {\loop
        \put(\xco,\yco){$\kcr$}
        \advance\yco by-2
        \ifnum\yco>-240
        \repeat
        \put(\xco,\yco){$\kbr$}}
        \put(-20,12){\ooo bacdefghidfghghdhededbihdgdfdfhhdheidhdhebaaahjhhdahba

hgdedge
   hgfdiehhgdigicba}
        \put(-20,-241.5){\ooo ababaighefdbfghgeahgdfgafagihdidihiidhiagfedhadbfd

ecdcdfa
   gdcbhaddhbgfchbgfdacfediacbabab}
        \end{picture}
        \par\vskip-8mm}
\def\headpic{                                           
        \indent
        \setlength{\unitlength}{.4mm}
        \thinlines
        \par
        \begin{picture}(29,16)
        \put(165,16){\line(1,0){4}}
        \put(170,16){\line(1,0){4}}
        \put(180,16){\line(1,0){4}}
        \put(175,0){\line(1,0){4}}
        \put(180,0){\line(1,0){4}}
        \put(185,0){\line(1,0){4}}
        \put(169,0){\line(0,1){16}}
        \put(170,0){\line(0,1){16}}
        \put(179,0){\line(0,1){16}}
        \put(180,0){\line(0,1){16}}
        \put(184,0){\line(0,1){16}}
        \put(185,0){\line(0,1){16}}
        \put(169,16){\oval(8,32)[bl]}
        \put(170,16){\oval(8,32)[br]}
        \put(179,0){\oval(8,32)[tl]}
        \put(185,0){\oval(8,32)[tr]}
        \end{picture}
        \par\vskip-6.5mm
        \thicklines}
\def\title#1#2#3#4{\border\headpic {\hbox to\hsize{#4 \hfill UMDEPP #3}}\par
        \begin{center} \vglue .5in {\large\bf #1}\\[.6in]
        {#2}\\[.1in] {\it Department of Physics and Astronomy}\\
        {\it University of Maryland, College Park, MD 20742}\\[1.5in]
        {\bf ABSTRACT}\\[.1in] \end{center} \begin{quotation}}  
\def\Title#1#2#3#4#5#6#7{\border\headpic
        {\hbox to\hsize{#7 \hfill UMDEPP #6}}\par
        \begin{center} \vglue .4in {\large\bf #1}\\[.4in]
        {#2}\\[.1in] {\it Department of Physics and Astronomy}\\
        {\it University of Maryland, College Park, MD 20742}\\[.1in]
        {#3}\\[.1in] {\it {#4}}\\ {\it {#5}}\\[.5in] {\bf ABSTRACT}\\[.1in]
        \end{center} \begin{quotation}}                 
\def\endtitle{\end{quotation}\newpage}                  

\def\ad{{\kern0.5pt
                   \alpha \kern-5.05pt \raise5.8pt\hbox{$\textstyle.$}\kern
0.5pt}}

\def\bd{{\kern0.5pt
                   \beta \kern-5.05pt \raise5.8pt\hbox{$\textstyle.$}\kern
0.5pt}}

\def\qd{{\kern0.5pt
                   q \kern-5.05pt \raise5.8pt\hbox{$\textstyle.$}\kern
0.5pt}}
\def\Dot#1{{\kern0.5pt
                   {#1} \kern-5.05pt \raise5.8pt\hbox{$\textstyle.$}\kern
0.5pt}}

\begin{document}

\def\gfrac#1#2{\frac {\scriptstyle{#1}}
        {\mbox{\raisebox{-.6ex}{$\scriptstyle{#2}$}}}}
\def\gg{{\hbox{\sc g}}}
\border\headpic {\hbox to\hsize{June 1996 \hfill {UMDEPP 96-99}}}
\par
\setlength{\oddsidemargin}{0.3in}
\setlength{\evensidemargin}{-0.3in}
\begin{center}
\vglue .08in
{\large\bf Why Auxiliary Fields Matter: The Strange\\
Case of the 4D, N = 1 Supersymmetric\\
QCD Effective Action (II) \footnote {Supported in part by National 
Science Foundation Grant PHY-91-19746 \newline ${~~~~~}$ and by NATO 
Grant CRG-93-0789}  }
\\[.72in]

S. James Gates, Jr.
\\[.02in]
{\it Department of Physics\\ 
University of Maryland\\ 
College Park, MD 20742-4111  USA}\\[.2in] 
{\bf {\tt gates@umdhep.umd.edu}}\\[2in]

{\bf ABSTRACT}\\[.002in]
\end{center}
\begin{quotation}
{Within a four dimensional N = 1 superspace, we present a new ansatz 
for the Skyrme term and for the {\underline {gauged}} WZNW term 
embedded into a superaction. We use the new chiral-nonminimal
(CNM) formulation for the effective low-energy action of 4D, N = 1 
supersymmetric QCD constructed by assigning right-handed components 
of Dirac fields to chiral multiplets and left-handed components of Dirac 
fields to nonminimal multiplets. It is noted that such a construction
likely allows for a new type of parity violation in low-energy 4D, N 
= 1 supersymmetric QCD.}  

\endtitle
\section{Introduction} 

~~~~The theory of supersymmetric effective actions has been found 
to possess a number of interesting surprises.  Over a decade ago, 
there first appeared in the literature \cite{0A} the class of 4D, 
N = 2 supersymmetric ``K\" ahlerian Vector Multiplet'' models 
\cite{A}\footnote{The locally supersymmetric version of these theories made
their first appearance in \cite{A1}.}.  This is a non-linear $\s$-model 
constructed from the 4D, N = 2 vector multiplet \cite{B} and a specific 
member of this class of models (the 4D, N = 2 effective action \cite{B0}) 
has recently been shown to possess information concerning the non-perturbative 
structure of the 4D, N = 2 supersymmetric Yang-Mills effective theory 
\cite{B1}.  This serendipitous circumstance has been the trigger of 
a huge amount of activity presently among theoretical physicists to 
study 4D, N = 2 supersymmetric effective actions (see \cite{BB} and
references therein). 

About a decade ago, there was also a period in which the topic of 4D,
N = 1 supersymmetric effective actions \cite{B2} was more actively 
pursued.  This prior epoch had as its trigger works on the structure of 
the QCD low-energy effective action.  The importance of the higher 
derivative terms such as the ``Skyrme'' term had been elucidated \cite{C0}
for soliton stability.  Another major advance occurred with the initial 
presentation of the ``topological'' approach \cite{CC} to the WZNW term 
\cite{D1}.  The works of reference \cite{B2} were attempts to describe 
supersymmetric extensions of various terms that appear in the low-energy 
QCD effective action, paying attention especially to terms involving spin-0 
fields in the non-supersymmetric limit of the effective action. There 
has also been some discussion of the 4D, N = 1 supersymmetric effective 
actions (for spin-1 fields \cite{DD}) that are related to the low-energy 
limit of open-string theory.

Yet from our prospective, throughout most of this period, there had
(until quite recently) remained one major interesting puzzle that we 
call the ``auxiliary freedom problem'' for 4D, N = 1 supersymmetric 
effective actions. To see this problem it suffices to consider a 4D, 
N = 1 supersymmetric theory involving only scalar multiplets where, 
however, the spin-0 components appear in an action with higher than 
second order derivatives and the spin-1/2 components appear with higher 
than first order derivatives.  This is the form of a generic term in 
an effective action.  We believe it is a desideratum to use 4D, N = 1 
superfields.  Thus accompanying the spin-0 and spin-1/2 fields there 
must be a set of auxiliary fields.  Since the physical fields' equations 
of motion are higher than usual order in derivatives, it is natural 
to expect that derivative operators appear in the equations of motion 
for auxiliary fields.  In this event, auxiliary fields become propagating!

In fact, this problem of propagating auxiliary fields in higher derivative
4D, N = 1 manifestly supersymmetric actions has been one of ``the dirty
little secrets'' bedeviling such theories but almost never discussed and
sometimes not even recognized.  For example, in their treatment of 4D,
N = 1 supersymmetric ``Skyrmions,'' Bergshoeff, Nepomechie and Schnitzer
(BNS work of \cite{B2}) noted that their ansatz for the form of 
the supersymmetric Skyrme term necessarily implies the presence of
propagating $F$-fields.  Similarly, although Nemeschansky and Rohm 
(NR work of \cite{B2}) did give an ansatz for the form of the
4D, N = 1 supersymmetric WZNW term, they did {\underline {not}} make 
mention of the fact that their proposal also necessarily contains 
propagating $F$-fields. We would be remiss, if we did not highlight 
the work of Karlhede, Lindstrom, Ro\v cek and Theodoridis \cite{DD} in 
this regard. To our knowledge, this has for a long time been the only 
work in the literature where the problem of propagating auxiliary fields 
in higher derivative 4D, N = 1 supersymmetric actions has been 
forthrightly addressed.

While for the effective action of 4D, N = 1 supersymmetric QCD we are 
not able to make a definitive argument that propagating auxiliary fields 
are ``bad,'' for compactified 4D, N = 1 superstrings and heterotic 
strings there is a potential for making such arguments for their low 
energy effective actions. The point in these theories is that the spectrum 
of the effective action is strictly controlled by string theory. Propagating 
auxiliary fields would have to correspond to higher mass ($m > 0$) modes 
of the string. If the spectrum of the string cannot accommodate the states 
described by the propagating auxiliary fields, that is reason to rule them 
out.

The present generally accepted proposal for the superfield description of 
the 4D, N = 1 supersymmetric low-energy QCD effective action has numbers 
of other problems even if one is willing to put aside the question of 
propagating auxiliary fields.  The previous work on 4D, N = 1 Skyrmions
\cite{B2} ends in part by concluding ``...both terms contain quartic time 
derivatives and lead to actions that are not bounded from below.'' 
Similarly, the authors find, ``In particular, one expects that the 
CP${}^1$ case ($\g =0$) should be special, corresponding to an infinitely 
thin rigid rod, having zero moment of inertia about the symmetry axis.  
This expectation is not bourne out by our explicit collective 
coordinate calculation.''  In the case of the previous work on the 
4D, N = 1 supersymmetric WZNW term, we find the conclusion
``...a Wess-Zumino term for this effective Lagrangian model of 
supersymmetric QCD exists and has the correct anomalies, although
we do not know a way to construct it explicitly.'' Furthermore,
the sigma model manifold associated with this older formulation
of the 4D, N = 1 supersymmetric WZNW term must necessarily possess
a non-compact geometry.

To our thinking, all of these were sufficiently severe problems 
to raise the question of whether there is something fundamentally 
lacking in our understanding of higher derivative 4D, N = 1 
superfield theory. By the end of our first effort to study this
class of theories \cite{ACG} we stated, ``In our opinion, more
work is required to reconcile supersymmetry with higher derivative
actions in four dimensions.''

Although our concerns along these lines have been constant and of
long duration, we have never had concrete alternative suggestions 
to those in the literature and which might avoid all the problems described 
above...until recently.   In one of our recent investigations \cite{SG}, 
we have found that there is a mechanism that permits the existence 
of a previously unknown class of higher derivative 4D, N = 1 manifestly 
supersymmetric actions involving propagating spin-0 and spin-1/2 fields 
and {\underline {no}} propagating auxiliary fields!  This result is the 
(0,1/2) multiplet analog of the (1,1/2) multiplet result \cite{DD}. Our 
mechanism for achieving this result uses 4D chirality in an essential 
way. We find that given a theory with Dirac particles, we can assign 
the right-handed spinor components to chiral multiplets \cite{G} and 
the left-handed spinor components to 4D, N = 1 nonminimal scalar 
multiplets \cite{C,D}. We call these chiral-nonminimal (CNM) models.
A CNM model, in and of itself, is not sufficient to suppress the 
propagation of auxiliary fields. To complete their suppression,
we impose strong\footnote{We define weak holomorphy as the condition that
all non-polynomial functions which determine \newline ${~~~\,~}$ actions
are holomorphic or chiral functions even though the actions are not 
supersymmetrically \newline ${~~~\,~}$ chiral. The 4D, N = 1 nonlinear 
$\s$-model involving chiral superfields and special K\" ahler geometry 
\newline ${~~~\,~}$ is an example of weak holomorphy.} holomorphy \cite{I},
the condition that the higher derivative action itself is holomorphic (i.e. 
supersymmetrically chiral).

It is the purpose of this present work to establish a new paradigm, the CNM 
\cite{SG} approach, to the description of 4D, N = 1 supersymmetric effective
actions for dynamical superfield multiplets of spins (0,1/2). We do {\underline
{not}} present derivations of our proposed CNM formulation from underlying
fundamental theories. Instead our emphasis is on the reconciliation that our
CNM approach provides between the 4D, N = 1 superfield formalism and the 
well-known characteristic structures of low-energy QCD. In a sense our 
discussion may be thought of as the introduction of a new phenomenological 
model of ``super-pion'' dynamics. The dynamical bosonic fields include the 
usual SU(3) pion octet, a second psuedo-scalar SU(3) octet (denoted by 
$\Theta$) isomorphic to the pion octet and a complex scalar SU(3) octet 
(denoted by $G_+$). The dynamical fermions consist of an SU(3) octet of 
Dirac particles (denoted by $\ell$).  Supersymmetry requires this spectrum 
in order to have equal numbers of bosons and fermions.  At this stage we 
defer to the future the issues of supersymmetry breaking, masses, etc. 
Instead in this work we will concentrate on sorting out various conceptual 
issues involved in our approach.

In chapter two we introduce the CNM description of scalar multiplet systems.
Chiral and nonminmal scalar multiplets are briefly reviewed.  The role of 
holomorphy is defined as essentially restricting the non-polynomial 
dependence of the effective action {\underline {solely}} to chiral 
superfields. The leading $\s$-model terms of the effective action are 
described.   Next the suggestion of \cite{SG} is reviewed in order to 
include higher derivative terms such as the Skyrme and WZNW terms. A 
simple proof is described that shows that no propagating auxiliary 
fields arise from the inclusion of the higher derivative sector. The 
chapter ends with setting up further investigation to study the 
uniqueness of the form of the higher derivative terms.

In chapter three we review the standard (N = 0) Skyrme and WZNW terms
and as well review the only works, known to us, in the literature where
an extension to include 4D, N = 1 supersymmetry had been attempted.
In particular, the auxiliary field sector of the work by Bergshoeff,
Nepomechie and Schnitzer on 4D, N = 1 skyrmions and as well the work
by Nemeschansky and Rohm on the 4D, N = 1 supersymmetric extension 
of the WZNW term are critically reviewed. It is shown (although no 
prior discussions along this line exist in the literature) that the 
auxiliary fields of the NR action are actually dynamical. We re-confirm
that similar behavior exists for the Skyrme term.

In chapter four the conceptual issues in regards to K\" ahler geometry
are confronted.  It is proposed that the nonminimal multiplets, unlike
chiral multiplets, are elements of the co-tangent bundle of the manifold
described by the chiral sub-sector of the complete CNM model. For the
first time it is noted that chiral superfield multiplets in supersymmetric
non-linear $\s$-models possess an intrinsic definition of spacetime 
chirality defined by the association of the spinors in the supermultiplets
with either the holomorphic or anti-holomorphic co-tangent bundles.
It is also noted that the spinors of nonminimal multiplets can be
used in addition to those of chiral multiplets to realize holomorphic
vector-like models with respect to the intrinsic definition.

In chapter five we discuss issues of irreducible super p-form
geometry and the CNM-WZNW action. The discussion begins with a 
review of the super $p$-form gauge supermultiplets as an exercise
to orient the reader regarding the structure of the simplex of
4D, N = 1 super irreducible p-forms. As another exercise, the 4D, 
N = 1 supersymmetric 4-form associated with the instanton
number density is reviewed. Next a complete supergeometry is
constructed for a super 4-form that contains the CNM-WZNW term.
It is found that such a super 4-form exists consistent with the
``p-form theorem'' of the 4D, N = 1 supersymmetric p-form simplex.
The CNM-WZNW action is shown to be the superspace integral of one 
component of the WZNW super 4-form similar to the fact that the 
superfield instanton index is one component of a super
4-form.

In chapter six we discuss, in a very preliminary way, some aspects
of the component fields described by the CNM approach. The
particle spectrum of dynamical fields is explicitly defined
from the corresponding superfields. Most amusingly, a mild
analogy with the structure of the Glashow-Salam-Weinberg
model is noted. This includes the introduction of a mixing
angle denote by $\g_{\rm S} $(analogous to the weak mixing angle) 
that is restricted to satisfy the condition $sin ( 2 \g_{\rm S})
\neq 0$.

In chapter seven we set out the issues and begin the task of
constructing the gauged version of the 4D, N = 1 supersymmetric
nonlinear $\s$-model, Skyrme and WZNW terms.  The usual (N = 0)
theory is written in such a way so as to facilitate an ansatz 
for its supersymmetric extension.  A proposal is made for the 
form of the supersymmetric ``pull-backs'' required to write the 
complete action.  

In chapter eight, we explore the implications of the CNM approach
to effective actions for a new re-formulation of its underlying
supersymmetric Yang-Mills gauge theory. It is shown that with
the use of nonminimal multiplets, there are two distinct definitions
of Dirac fields within the context of supersymmetric theories.
For one version with gauge group $G$ in the WZ gauge (that used 
exclusively in the literature), the gauge group outside of the 
WZ gauge is found to be $G_V \otimes G_A$.  For the CNM version 
with gauge group $G$ in the WZ gauge, the gauge group is found to 
be  $G_c$, the complexified extension.

In our conclusions, we discuss the possible significance of our
new results for a disagreement on the structure of the low-energy
effective action of heterotic string theory. We also note that a
CNM model for the effective action seems to characteristically predict
the {\underline {breaking}} of parity! Thus, the model is one where
P-violation occurs even for the strong interactions.  Two appendices
containing presentations of related results are included.

\section{4D Chirality as the Solution to the Auxiliary Freedom Problem}

~~~~The underlying fermions, the quarks, of the fundamental QCD theory
are Dirac particles. Using the chirality projectors, we can always
split a Dirac field into right-handed and left-handed components.
On the other hand, the fundamental fermions contained in 4D, N = 1 
superfields may be considered as either Majorana or Weyl particles. 
Thus, to embed any theory of Dirac particles into a supersymmetric
theory, one is faced with an initial choice of to what supersymmetric
representation should the two 4D chiral components of the Dirac field
be assigned.  Any brief consultation of the huge body of the literature 
shows that the state-of-the-art in constructing phenomenologically 
relevant 4D, N = 1 supersymmetric field theories has been to assign 
both the right-handed components as well as the charge conjugates of 
the left-handed components to chiral multiplets \cite{G}. If this was
the only off-shell (0, 1/2) supersymmetric representation we would 
be forced to do this.

As was pointed out a long time ago \cite{C,D}, this is not the case.
There exist a number of other off-shell (0, 1/2) supersymmetric 
representations. For example, there is the linear multiplet \cite{SSG}
containing the axion and dilaton that occur in 4D, N = 1 superspace 
geometry \cite{GMOV} arising as a limit of heterotic string theory.
However, among all the variant representations to the chiral multiplet
there is one, the nonminimal multiplet, which is unique. The nonminimal
multiplet (and its infinite family of daughters) is the only variant 
to the chiral multiplet which does not contain component gauge fields.
This singles this 4D, N = 1 supersymmetric representation out as
being eminently suited to play the role of a matter multiplet in
phenomenologically interesting proposals.

We can compare the fields of the two different multiplets by looking at the
following table.
\begin{center}
\renewcommand\arraystretch{1.2}
\begin{tabular}{|c|c| }\hline
${\rm Phys.~vs.~Aux.}$
& ${\rm~~\,P~~~~P~~~~A~~~~~A~~~~\,A~~~~}$  \\ \hline
${\rm Eng.~Dim.}$
& $ ~\,1 ~~~~~ {\frac 32}~~~~~{\frac 32}~~~~~~2~~~~~{\frac 52}~~~ $  \\ \hline
\hline
$  {\rm {Chiral~SF}} $ &  $ A~~~~ \psi_{\a} ~~~{~}~~~~~~~~ F ~~~~~~~~~
$ \\ \hline
$  {\rm {Nonminimal~SF}} $ &  $ B~~~ {\Bar \zeta}_{\Dot \a}~~~
{\r}_{\a} ~~~ H, \, p_{\underline a} ~~~ {\Bar \b}_{\ad}
 $ \\ \hline
\end{tabular}
\end{center}
\centerline{{\bf Table I}}
Chiral versus nonminimal superfields are defined by the respective
conditions,
$$
{\Bar D}_{\Dot \a} \Phi = 0 ~~~,~~~ {\Bar D}^2 \S = 0 ~~~,
\eqno(2.1) $$
which in a real sense may be called the ``Bianchi Identities''
for each multiplet.  The simplest actions for describing 
the dynamics of such superfields are,
$$
{\cal S}_{WZ} ~=~ \int d^4 x d^2 \q \, d^2 {\Bar \q} ~ {\Bar \Phi} \, \Phi
~~~~,~~~
{\cal S}_{NM} ~=~ -~ \int d^4 x d^2 \q \, d^2 {\Bar \q} ~ {\Bar \S} \, \S 
~~~,
\eqno(2.2)
$$
where the explicit component forms for these (as well as the definition
of the components) can be found in our earlier paper \cite{SG}. The
equations of motion that follow from these actions imply that $A$ and
$B$ satisfy massless Klein-Gordon equations, $\psi_{\a}$ and ${\Bar 
\zeta}_{\ad}$ satisfy massless Dirac equations and that all remaining
fields vanish. The superfield equations of motion that follow from these 
actions (2.2) as well as the defining conditions in (2.1) above reveal 
that there exists a type of duality between these two multiplets
(see {\it {Superspace}} \cite{SP}).  Let us here define ``Poincar\' e 
dual pairs.'' We will call two fields (or superfields) Poincar\' e dual 
pairs if when we exchange the Bianchi identity with the equation of 
motion for one member of the pair, we arrive at the other member.
We'll describe pairs of fields with this property as possessing
``Poincar\' e duality.'' It can be seen that this definition is
nothing but a generalization of the electric-magnetic duality of
the photon. So for example, the sum of the two actions in (2.2) can
be said to possess ``Poincar\' e duality invariance'' since it would
be meaningless to speak of electric-magnetic duality invariance
in  a theory with no dynamical spin-1 field.  As well it is worth 
mentioning that the sum also realizes on-shell N = 2 supersymmetry.
\begin{center}
\renewcommand\arraystretch{1.2}
\begin{tabular}{|c|c|c| }\hline
${~}$ & ${\rm Constraint}$ & ${\rm Equation~of~Motion}$  \\ \hline \hline
${\rm {E.~\&~M.}}$ & $  d\, F  = 0
$ & $d{}^* F = 0$  \\ \hline
${\rm {Chiral~SF}}$ & $  {\Bar D}_{\Dot \a} \Phi = 0
$ & $ D^2 \Phi = 0$  \\ \hline
${\rm {Nonminimal~SF}}$ & $D^2 {\Bar \S} = 0$ & $ {\Bar D}_{\Dot a} 
{\Bar \S} = 0$  
\\ \hline
\end{tabular}
\end{center}
\centerline{{\bf Table II}}
Thus, we finally see that our proposal to embed the two different
chiral components of a Dirac spinor into chiral and nonminimal
superfields is equivalent to imposing the condition that only 
Poincar\' e dual pairs $(\Phi, \, \S)$ should be thought of as the 
4D, N = 1 supersymmetric definition of a Dirac spinor.

Although we began our discussion considering the underlying Dirac
fields of QCD, it is ultimately only the low-energy 4D, N = 1 
supersymmetric QCD effective action in which we wish to implement 
our proposal of the use of dual pairs. As such, we wish to embed 
the pion octet into a supersymmetric formulation. Due to 
supersymmetry, these are also accompanied by their superpartners, 
the pionini octet, as well as other spin-0 fields. So we introduce 
Poincar\' e dual pairs ($\Phi^{\rm I} , \,\S^{\rm I}$) where ${\rm 
I} = 1,..., 8$. The most general non-linear $\s$-model term 
involving these superfields takes the form,
$$
{\cal S}_{\s} ~=~  \int d^4 x d^2 \q d^2 {\Bar \q} ~ {\Hat \O} 
(\, \Phi, {\Bar \Phi} ;  \S, {\Bar \S} \, ) ~~~~,
\eqno(2.3) $$
where ${\Hat \O}$ is a K\" ahler potential. In order to obtain a 
slightly more explicit form of the K\" ahler potential, we assume
three constraints; (a.) the flat limit ought to correspond to the 
free action for these multiplets, (b.) ${\Hat \O}$ is subject to 
weak holomorphy and (c.) the action is no more than quadratic in the 
nonminimal multiplet. These conditions are sufficient to imply that 
the K\" ahler potential has the form
$$
{\Hat \O} ~=~ \frac 12 \, \Big[ ~  [~ {\Bar \Phi}^{\rm I} {\rm H
}_{\rm I}({\Phi}) ~+~ {\Bar \S}{}^{\rm I} {\rm E}_{\rm I} ({\Phi})
~-~ {\cal J}_{\rm {I\, J}}(\Phi) {\Bar \S}{}^{\rm I} 
{\S}^{\rm J}  ~+~ (\, {\rm H}_{\rm {I\, J}}(\Phi)  
 ~+~ {\Bar {\rm K}}_{\rm {I\, J}} ({\Bar \Phi})  \, ) {\S}^{\rm I} 
{\S}^{\rm J} ] ~+~ {\rm {h.\, c.}} ~ \Big] ~~~~,
\eqno(2.4) $$
where ${\rm E}_{\rm I}(\Phi)$, ${\rm H}_{\rm I}(\Phi)$, ${\rm H}_{{\rm 
I}\, {\rm J}} (\Phi)$, ${\rm K}_{{\rm I}\, {\rm J}} (\Phi)$ and ${\cal 
J}_{\rm {I\, J}}(\Phi)$ are holomorphic. It can be seen that one 
special choice of these is given by 
$$ 
{\rm H}_{\rm I} ~=~ \pa_{\rm I} {\rm H} ~~~,~~~ {\cal J}_{\rm {I\, J}}
~=~ \frac 12 \pa_{\rm I} \pa_{\rm J} {\rm H} ~~~, ~~~ {\rm E}_{\rm 
I} ~=~ {\rm K}_{\rm {I\, J}} ~=~ {\rm H}_{\rm {I\, J}} ~=~ 
0 ~~~,
\eqno(2.5)
$$
for which the equation $\pa_{\rm I} {\rm H}_{\rm J} = 2 {\cal J}_{\rm 
{I\, J}}$ is satisfied. The results in (2.4) and (2.5) bare a 
strikingly similar appearance to K\" ahlerian vector multiplet 
models. In fact, it seems extremely likely that these two classes 
of models are related to each other via the RADIO technique \cite{GL}. 
Finally, we will argue that the natural geometric interpretation of 
the Poincar\' e dual pairs $(\Phi, \, \S)$ is that the coordinates of 
a complex manifold ${\cal M}$ are provided by $\Phi$ while $\S$ are 
elements of ${}^* T_p({\cal M})$, the dual to the {\it {holomorphic}} 
vector bundle of the manifold. Since this together with (2.4) and (2.5) 
obviously defines a very particular type of fibered K\" ahler geometry, 
we will call this ``specular K\" ahler geometry.''

The action of equation (2.3) describes a non-linear $\s$-model and as such 
no higher order derivative terms are present.  It is well known that such
higher order derivative terms are present in the standard QCD low-energy
effective action. So in order to introduce such terms here we must go
beyond (2.3).  As we noted in \cite{SG} if we assume that strong holomorphy
is satisfied, then the form of the higher derivative terms is completely
determined
$$
{\cal S}_{{\rm H.\, D.}} ~=~  (\g')^{3} \Big[ \int d^4 x \, d^2 \q ~ 
\sum_{N = 4} {\cal L}_{(N)} ~+~ {\rm {h. \, c.}}  ~ \Big] ~=~  (\g')^{3} 
\Big[~ {\cal S}_{{\rm H.\, D.}}^{(4)} ~+~{\cal S}_{{\rm H.\, D.}}^{(6)}
~+~...  ~ \Big]  ~~~~,
\eqno(2.6) $$
where we introduced an expansion parameter\footnote{One choice 
is the QCD cutoff, $\L_{QCD}$.  Another is given by $f_{\pi}$.
} of appropriate dimensions denoted by $\g'$.  
Thus, we demand that all terms in (2.6) are determined by a set of 
holomorphic tensors ${\cal J}_{\rm I_1  \, J_1 \, ... \,  K_1 \, ...}^{A 
\, k_j} (\Phi)$. Geometrically these are to be thought of as proper
holomorphic tensors defined over the various bundles of the K\" ahler 
manifold.
The $A$ label denotes the different irreducible 
representations possible for a fixed number of K\" ahler manifold
indices and integers $k_1, \, ... \, k_N$ denote other ``naming'' labels.
We further introduce $ P^{k_i \, {\underline c}_{1} \, ... }_{A \, 
{\ad}_1 \, {\Dot \b}_1 \, ... }$ as a set of constant tensors 
carrying non-trivial Lorentz representations. Utilizing the 
${\cal J}$-symbols and $P$-symbols, we write the ${\cal L}_{(N)}$'s 
in the forms
$$\eqalign{ {~~~~~~~~~~~~~~~}
{\cal L}_{(N)} &\equiv~ \sum P^{k_i \, {\underline c}_{1}  \, ...  
}_{A \, {\ad}_1 \, {\Dot \b}_1 \, ... } {\cal J}_{\rm I_1 
\, J_1 \, ... \,  K_1 \, ...}^{A \, k_i} (\Phi) (  \prod 
{\cal G}^{ {\rm I}_i \,{\rm J}_i ~ {\ad}_i \, {\bd}_i} ) ( \prod
{\Hat {\cal G}}^{ {\rm K}_j}_{{\underline c_1} ... {\underline 
c}_{k_j}}) ~~~~, \cr 
{\cal G}^{ {\rm I}_i \,{\rm J}_i ~ {\ad}_i \, {\bd}_i} &\equiv~ (\g')^{- 3} 
( {\Bar D}^{{\ad}_i} \S^{{\rm I}_i} \,) ( {\Bar D}^{{\bd}_i} 
\S^{{\rm J}_i} \, )  ~~~~, \cr
~~~ {\Hat {\cal G}}^{ {\rm K}_j}_{{\underline c_1} ... {\underline c}_{k_j}} 
&\equiv~ (\g')^{- (1 + k_j)}
( \pa_{{\underline c}_1} {}_{{~}^{...}} ~ \pa_{{\underline c}_{k_j}} 
\Phi^{{\rm K}_j} \,) ~~~~,}
 \eqno(2.7) $$
where $N$ denotes the number of bosonic fields with derivative operators 
that appear in  ${\cal L}_{(N)}$ (i.e. $N$ equals twice the number of 
factors of ${\cal G}$ plus the number of factors of ${\Hat {\cal G}}$ in a 
given term). It can be seen that if ${\cal G}$ appears other than linearly, 
these actions possess no purely bosonic terms.  Thus to have terms with the
property that they are non-trivial in the purely bosonic limit, we need 
only keep ${\cal G}$ to the first power.

The proof that (2.7) contains no propagating auxiliary field is very 
simple. As a first step let us concentrate only on possible bosonic
terms.  First, since it is a chiral superfield (${\Bar D}_{\ad} 
{\cal L}_{(N)} = 0$) we only need to evaluate it by applying $D^{\a} 
D_{\a}$ to obtain component results. Next we have to perform the 
differentiations. If both $D$'s act on the ${\cal J}$-term, the most 
we can get is an $F$ auxiliary field.  When evaluated at $\q = 0$, 
the remaining terms are purely physical fields. Similarly, if both 
$D$'s act on ${\cal G}$, the most we can get are terms quadratic in 
the $p_{\un a}$ auxiliary field. Finally if both $D$'s act on ${\Hat 
{\cal G}}$, the most we can get is an $F$ auxiliary field which 
however has derivatives acting upon it. This term is apparently 
dangerous until we realize that all spacetime derivatives may be 
integrated ``off'' of the $F$ auxiliary field and onto the remaining 
factors which are themselves physical fields. The extension to 
include fermionic terms is a straight forward exercise.

Now we come the the central suggestion of our new formulation of
the low-energy 4D, N = 1 supersymmetric QCD effective action.
We propose that it should be written as 
$$
{\cal S}_{\rm eff}^{{\rm SUSY}}(QCD) ~=~ 
{\cal S}_{\s} ~+~ {\cal S}_{{\rm H.\, D.}} ~~~.
\eqno(2.8) $$
This action has two important properties. Holomorphy is completely manifest 
because the nonlinear functions ${\rm E}_{\rm I}$, ${\rm H}_{\rm I}$, 
${\cal J}_{{\rm I}\, {\rm J}}$, ${\rm H}_{{\rm I}\, {\rm J}}$ and 
${\cal J}_{\rm I_1  \, J_1 \, ... \,  K_1 \, ...}^{A \, k_j}$ determine 
the explicit form of the action.  The equations of motion for {\underline 
{all}} auxiliary fields are {\it {algebraic}} so that no auxiliary fields 
propagate.

However, the use of this mechanism for suppression of propagating auxiliary
fields comes at a price.  The roles of the right-handed 
$\psi$-spinors (contained in the chiral superfields) are completely different 
from that of the left-handed $\zeta$-spinors (contained in the nonminimal 
superfields).  This should be particularly obvious by noting that the 
factors of ${\Bar D \S}$ in the higher derivative expansion correspond 
to the $\z$-spinors.  Supersymmetry itself forbids the symmetrical 
appearance of the different types of spinors in (2.7).   Thus, there is 
a possibility to realize a breaking of parity in this proposed 4D, N = 
1 supersymmetric extension of the QCD effective action.

We end this chapter with a few comments on research that still needs
to be undertaken to completely clarify the issue of auxiliary-free
higher derivative terms in theories involving 4D, N = 1 scalar
multiplets. The first question that comes to mind is whether
there are other auxiliary-free higher derivative terms that
{\underline {cannot}} be expressed in terms of chiral superfield Lagrangians.
Along these lines, there are two classes of actions that will
be studied in the future. We represent these two classes in
the form of two actions
$$
{\cal S}_{{\rm Class-I}} ~=~  \int d^4 x \, d^2 \q ~ d^2 {\Bar \q} ~ 
\Big[ \, ( {\Bar D}^{\ad} \S^{\rm I} \,) ~ (D^{\a} {\Bar \S}^{\rm J} 
\,) ~ {\cal Y}_{ {\rm I} \,{\rm J}\, {\un a} \,}^{\rm Class-I} ~\Big]  
~~~ ,
\eqno(2.9) $$
and as well
$$
{\cal S}_{{ Class-II}} ~=~  \int d^4 x \, d^2 \q ~ d^2 {\Bar \q} ~ 
\Big[ \, {\cal G}^{ {\rm I} \,{\rm J} ~ {\ad} \, {\bd}} \, 
{\Bar {\cal G}}^{ {\rm K} \,{\rm L} ~ {\a} \, {\b}} ~
{\cal Y}_{ {\rm I} \,{\rm J}\, {\rm K} \,{\rm L} ~ {\un a} \, 
{\un b} }^{Class-II} ~\Big]  ~~~ ,
\eqno(2.10) $$
where the ${\cal Y}$'s are, in complete generality, functions of 
$\Phi$ and $\S$ and any of their derivatives (either bosonic
or fermionic).  We point out that the split above is
somewhat artificial since in fact ${\cal S}_{{ Class-II}}$ 
is a special case of ${\cal S}_{{ Class-I}}$. However, this
split is useful once we note (2.10) together with choosing
the function ${\cal Y}_{ {\rm I} \,{\rm J}\, {\rm K} \,{\rm L} 
~ {\un a} \, {\un b} }^{Class-II}$ according to (where ${m_i}$
are integers)
$$
{\cal Y}_{ {\rm I} \,{\rm J}\, {\rm K} \,{\rm L} ~ {\un a} \, 
{\un b} }^{Class-II} ~=~ {\cal Y}_{ {\rm I} \,{\rm J}\, {\rm K} 
\,{\rm L} ~ {\un a} \, {\un b} }^{Class-II} (\Phi, \S,
{\Bar \Phi} , {\Bar \S}, \pa^{m_1} \Phi, \pa^{m_2} \S, 
\pa^{m_3} {\Bar \Phi} ,  \pa^{m_4} {\Bar \S} \,) ~~~,
\eqno(2.11) $$ 
{\underline {necessarily}} implies purely bosonic higher derivative 
terms in the component level action.  What remains, however, is to 
find the conditions required on ${\cal Y}_{ {\rm I} \,{\rm J}\, {\rm 
K} \, {\rm L} ~ {\un a} \, {\un b} }^{Class-II}$ to insure auxiliary
freedom in even in its fermionic sector.  A general analysis will
carried out in the future.  

Finally, we end by noting that the mechanism that we use to construct 
4D, N = 1 supersymmetric auxiliary-free higher derivative actions 
for spin (0.1/2) multiplets bares some resemblance to that used in
the (1, 1/2) case \cite{DD}.

\section{Review of N = 0 and `Old' N = 1 Supersymmetric
Skyrme and WZNW Terms}

~~~~The standard Skyrme term has the familiar form
$$ \eqalign{ 
{\cal S}_{Skyrme} &=~   \frac 1{32 e^2  C_2} \, \int d^4 x ~
{\rm {Tr}} [\, ( \pa^{[ \, \underline a} U \,) ~ (\pa^{\underline b 
\, ] } U ^{-1} \,) ( \pa_{[ \, \underline a} U \,) ~ (\pa_{\underline b 
\, ] } U ^{-1} \,) \,] \cr
&=~ \int d^4 x ~ \, k_{m \, n \, r \, s} \, (\Pi) ~ ( 
\pa^{[ \, \underline a} \Pi^m \,) ~ (\pa^{\underline b \, ] } \Pi^n \,) ~
(\pa_{\underline a} \Pi^r \,) ~ (\pa_{\underline b} \Pi^s \,) ~~~,}
\eqno(3.1) $$
where we use the notation of appendix A in reference \cite{SG}.
On the other hand, the WZNW term can be expressed as
$$
{\cal S}_{WZNW} ~=~ \int d^4 x \, \e^{{\underline a}{\underline b}
{\underline c} {\underline d}}{\cal J}_{m \, n \, r \, s} (\Pi) 
(\pa_{\underline a} \Pi^m \,) \, (\pa_{\underline b} \Pi^n \,) \,
(\pa_{\underline c} \Pi^r \,)  \, (\pa_{\underline d} \Pi^s \,) ~~~~. 
\eqno(3.2) $$
A long time ago we \cite{ACG} gave a local set of geometric conditions 
which seemed to us to be necessary but not sufficient for these actions
to correspond to the Skyrme and WZNW terms respectively.  We may
regard the fields $\Pi^m$ as the coordinates of some manifold
endowed with a metric $g_{m \, n}$. We further require that this
metric possess a set of Killing vectors $\xi^m$ such that Killings
equation takes the form ${\cal L}_\xi (g) ~=~ 0$ expressed in terms
of Lie differentiation. Necessary conditions for the local tensors 
$k_{m \, n \, r \, s}$ and ${\cal J}_{m \, n \, r \, s}$ to generate 
Skyrme and WZNW terms via the actions above are;
$$ 
{\cal L}_\xi (k) ~=~ 0 ~~~~,~~~~ d {\cal L}_\xi ({\cal J}) ~=~ 0
~~~, 
\eqno(3.3) $$
where $d$ in the second equation denotes the exterior derivative
with respect to the $\Pi$-coordinates. Note that neither Lie nor 
exterior differentiation requires a Riemannian geometry.  It is 
also obvious that the Skyrme term and WZNW terms are related by 
Poincar\' e duality.

Prior to examining the form of our new supersymmetric proposal, we feel 
that it is useful to review the prior results. We wish to explicit 
demonstrate the reasons for our long standing concerns about the old
proposals .  Namely in both of these suggestions, the auxiliary 
$F$-field of the chiral multiplets possess equations of motion that 
are {\underline{not}} purely algebraic.

Bergshoeff, Nepomechie and Schnitzer \cite{B2} (BNS) have made, to 
our knowledge, the only explicit proposal for the form of the 4D, N = 
1 supersymmetric Skyrme term given as;
$$ \eqalign{ {~~~~~}
{\cal S}_{Skyrme}^{BNS} ~=~  \int d^4 x d^2 \q  d^2 {\Bar \q} ~ \Big\{ \, 
\a \, \Big[ ~&( {\Bar \nabla}^{\Dot \a} {\Bar \Phi}^i \,) 
( {\Bar \nabla}_{\Dot \a} {\Bar \Phi}^j \,) ( {\nabla}^{\a} {\Phi}_i \,) 
( {\nabla}_{\a} {\Phi}_j \,) ~\Big] ~+~ \cr 
\b \Big[ ~  &( {\Bar \nabla}^{\Dot \a} {\Bar \nabla}_{\Dot \a} 
{\Bar \Phi}^i \,) ( {\nabla}^{\a} {\nabla}_{\a} {\Phi}_i 
\,) ~\Big]  ~\Big\} ~~~, }
\eqno(3.4) $$
where this action is for a 4D, N = 1 supersymmetric CP${}^1$ model
in particular and we have neglected the normalization since it
is irrelevant to our discussion. We wish to solely concentrate upon
how the auxiliary fields appear in this action. (The remaining
purely bosonic terms can be found in their work.) It is a straight
forward calculation to show that $F$-field dependent terms are
given by
$$   \eqalign{ 
{\cal S}_{Skyrme}^{BNS} ~=~  \int d^4 x ~  \Big\{ \, ...- \a \, 
\Big[ \,\, & ~ {\Bar F}^i F_j (\nabla^{\un a} A_i \,)  (\nabla_{\un a} 
{\Bar A}^j \,) + ~ {\bar F}^i F_i (\nabla^{\un a} A_j \,)  (\nabla_{\un 
a} {\Bar A}^j \,) \cr 
&+ ~ 4 {\bar F}^i F_j {\bar F}^j F_i ~\Big]  ~+~ \b \Big[ ~ 
{\Bar F}^i \, (\nabla^{\un a} \nabla_{\un a} F_i \,)   ~-~  F_i 
( D')^i {}_j {\Bar F}^j \cr
& +~ {\Tilde {\Bar F}}^i {\Tilde {F}}_i ~ \Big] ~ \Big\} ~~,
} \eqno(3.5) $$
where the following definitions are to be used,
$$  {\Tilde {F}}^i ~\equiv~ (\nabla^{\un a} \nabla_{\un a} {\Bar A}^i \,) 
~-~ [ \, ( D') \, , \, {\Bar A} \, ]^i ~~~,~~~ ( D')^i {}_j  ~\equiv~ 
(\nabla^{\un a} A_j \,)  (\nabla_{\un a} {\Bar A}^i \,)  ~-~ 
{\Bar F}^i F_j  ~~~.
\eqno(3.6) $$
In evaluating the last two terms of (3.5) we only retain the $F$-dependent
pieces.  Some very interesting features can be see here.  Foremost
for nonvanishing $\b$, the $F$-field acquires a mass proportional to $\b^{-
\frac 12}$.  While it is true that for $\b = 0$, the $F$-field again 
becomes non-propagating, it is known in the BNS 4D, N = 1 extension 
of the Skyrme term that for $\b = 0 $, the Skyrmion is not stabilized.  

Nemenschansky and Rohm \cite{B2} (NR) have proposed that the 4D, N = 1 
supersymmetric extension of the WZNW term is of the form
$${\cal S}_{WZNW}^{NR} ~=~  \int d^4 x d^2 \q  d^2 {\Bar \q} ~ \Big[ 
\, \b_{\rm { I \,J  {\bar {\rm K}}}}(\, \Phi, \, {\Bar \Phi})  ( D^{\a} 
\Phi^{\rm I}\,) ( \pa_{\a \bd} \Phi^{\rm J} \,) ( {\Bar D}^{\bd} {\Bar 
\Phi}^{\bar {\rm K}} \,) ~+~ {\rm h.} {\rm c.}  ~\Big] ~~~,
\eqno(3.7) $$
where $\b_{\rm { I \,J  {\bar {\rm K}}}}$ is not holomorphic.  If we
impose the condition of weak holomorphy, then this expression must be
modified to
$$
{\cal S}_{WZNW}^{mod.\, NR} ~=~  \int d^4 x d^2 \q  d^2 {\Bar \q} ~ \Big[ \, 
{\cal J}_{\rm { I \,J  {\bar {\rm K}} \, {\bar {\rm L}} }}(\, \Phi)  
( D^{\a} \Phi^{\rm I}\,) ( \pa_{\a \bd} \Phi^{\rm J} \,) ( {\Bar D}^{\bd} 
{\Bar \Phi}^{\bar {\rm K}} \,) \, {\Bar \Phi}^{\bar {\rm L}} ~+~ {\rm h.} 
{\rm c.}  ~\Big] ~~~,
\eqno(3.8) 
$$
which looks very similar to our result in ref. \cite{SG}. This eliminates
all 6-fermion terms also. However, even this modification has other 
difficulties within the structure of K\" ahler geometry and the realization
of isometries. Furthermore, the problem described immediately below remains
even with this modification.

Returning now to (3.7), a large number of the component level terms contained 
in this action were presented before \cite{B2}. However, the critical (for 
our purposes) auxiliary field terms, denoted by ${\cal L}_{aux.}$, were not 
explicitly given. Again it is a straight forward calculation to show that 
$F$-field dependent terms are given by
$$ \eqalign{
{\cal S}_{WZNW}^{NR} ~=~  \int d^4 x \Big[ ~ &... ~-i 4 (\, 
\b_{\rm { I \,J  {\bar {\rm K}}}}  \,-\, \b_{\rm { J \,I  {\bar 
{\rm K}}}} \, ) ( \pa^{\un a} A^{\rm I} \,) 
( \pa^{\un a} { F}^{{\bar {\rm J}}} \,) {\Bar F}^{{\bar {\rm K}}} 
~+~ {\rm h.} {\rm c.}   \cr
&~~+~ i 4 (\, \b_{\rm { I \,J  {\bar {\rm K}} ,\, {\bar {\rm L}}}}
\,-\, \b_{\rm { I \,J  {\bar {\rm L}} ,\, {\bar {\rm K}}}} \, ) 
\, F^{\rm I} ( \pa^{\un a} A^{\rm J} \,) \, {\Bar F}^{\bar {\rm K}}
( \pa^{\un a} {\Bar A}^{{\bar {\rm L}}} \,) ~+~ {\rm h.} 
{\rm c.} ~~ \Big] ~~~.
} \eqno(3.9) $$
We see that for the NR 4D, N = 1 extension of the WZNW term, the 
propagation of the $F$-fields proceeds only by ``non-linear $\s$-model 
mixing.''  Thus, equation (3.9) above explicitly demonstrates 
the propagation of $F$-fields. Thus, barring miraculous accidents, 
the $F$-fields become dynamical in NR WZNW term.

In closing this chapter, we note the generality of these results, 
although we have been investigating within the confines of the 4D, N = 1 
supersymmetric low-energy effective QCD action, our comments apply to any 
higher derivative manifestly 4D, N = 1 supersymmetric action (as all SUSY 
effective actions are) whether the application is effective SUSY QCD,  
MSSM or even SUSY haplon or preon type models. 

\section{K\" ahler Geometric Interpretation}

~~~~At the time we made our introduction of this new approach it was not 
completely clear what geometric structure could undergird our proposal.
In this section we would like to suggest that (as was alluded to 
in section two), the most natural geometric structure seems to be
provided by a K\" ahler manifold together with its holomorphic co-tangent 
bundle.  Let us explore how this seems consistent.

We first wish to establish a notation for the various geometrical
structures associated with a K\" ahler manifold denoted
by ${\cal M}$.  Let $T_p({\cal M})$ denote the holomorphic vector 
bundle (tangent plane) at point $p$.  Also let ${\Bar T}_p({\cal M})$ 
denote the anti-holomorphic vector bundle at point $p$.  Additionally 
let ${}^* T_p({\cal M})$ denote the holomorphic co-tangent bundle (dual 
to $T_p({\cal M})$).  Finally let ${}^* {\Bar T}_p({\cal M})$ denote the
anti-holomorphic co-tangent bundle (dual to $ {\Bar T}_p({\cal M})$). 

Having established a notation for the various K\" ahler geometrical 
objects of interest, we observe that the usual 4D, N =  1 supersymmetric 
non-linear $\s$-model implies that its various fields are among 
the elements of the various geometrical structures according to,
$$ \eqalign{
&(d A , \, \psi_{\a} \,)  ~\in ~ {}^* T_p({\cal M}) ~~~,~~~
 (d {\Bar A} , \, {\Bar \psi}_{\Dot \a} \,) ~ \in ~ {}^* {\Bar T}_p(
{\cal M})  ~~~, \cr
&{~~~~}( A, \, {\Bar A} ) ~ \in ~  {\cal M} ~~~, ~~~ \pa / \pa A
 ~ \in ~  T_p({\cal M}) ~~~,~~~ \pa / \pa {\Bar A}
 ~ \in ~ {\Bar T}_p({\cal M}) ~~~.}
\eqno(4.1) $$
These equations allow us to define a {\it {holomorphic}} chirality 
for a theory with chiral multiplets which appear in a non-linear 
$\s$-model.  Namely, we see that all undotted (``right-handed'')
spinors are elements of ${}^* T_p({\cal M})$ and all dotted 
(``left-handed'') spinors are elements of ${}^* {\Bar T}_p({\cal M})$. 
Thus, by holomorphic chirality we mean that there exists a specific 
correlation between the space-time chirality of the spinors and the 
co-tangent bundles (i.e. no undotted spinors are elements of 
${}^* T_p({\cal M})$ and vice-versa).

Since $\Phi^{\rm I}$ are the coordinates of a K\" ahler manifold,
there must exist some K\" ahler metric. Let us assume that the K\" ahler 
metric which we denote by $g_{{\rm I} \, {\Bar{\rm K}}}$ possesses a 
set of isometries generated by the holomorphic co-tangent fields 
$\xi^{\rm I}(\Phi)$ which thus satisfy Killings equation,
$$ 
{\cal L}_{\xi} (g_{{\rm I} \, {\Bar{\rm K}}}) ~=~ \xi^{\rm L}
\pa_{\rm L} g_{{\rm I} \, {\Bar{\rm K}}} ~+~ {\Bar \xi}^{\Bar {\rm L}}
{\Bar \pa}_{{\Bar \rm L}} g_{{\rm I} \, {\Bar{\rm K}}} ~+~( \pa_{\rm I} 
\xi^{\rm L}) g_{{\rm L} \, {\Bar{\rm K}}} ~+~ ({\Bar \pa}_{\Bar {\rm K}} 
\, {\Bar \xi}^{\Bar {\rm L}}) g_{{\rm I} \, {\Bar{\rm L}}} ~=~ 0 ~~~. 
\eqno(4.2) $$

Since the chiral superfields are the coordinates for the K\" ahler 
manifold, under the action of the isometries these transform according 
to the rule
$$ \d_{\xi} \Phi^{\rm I} ~=~ \xi^{\rm I} (\Phi) ~=~ \a^{(A)} 
\xi^{\rm I}_{(A)} (\Phi)  ~~~.
\eqno(4.3) $$
In the second part of the equation above we have noted that in general 
there will be numbers of independent such isometries ($(A)$ = 1, ...,
$m$). For each such isometry, we can introduce a constant parameter
$\a^{(A)}$.

On the other hand, the nonminimal superfields may be identified as elements 
of ${}^* {T}_p({\cal M})$. Thus, their transformation laws are determined 
to be,
$$ \d_{\xi} \S^{\rm I} ~=~ ( \pa_{\rm K} \xi^{\rm I} ) \S^{\rm K}
~=~ \a^{(A)} ( \pa_{\rm K} \xi^{\rm I}_{(A)} ) \S^{\rm K} ~~~. \eqno(4.4) $$
Equations (4.3) and (4.4) also give us a very deep insight into why
our chiral-nonminimal (CNM) superfield models are fundamentally different from 
the usual chiral superfield models. The respective transformation laws 
of the component spinors contained in each multiplet are
$$
\d_{\xi} \psi_{\a}^{\rm I} ~=~ ( \pa_{\rm K} \xi^{\rm I} ) \, \psi_{\a}^{\rm K}
~~~,~~~ 
\d_{\xi} {\Bar \z}_{\ad}^{\rm I} ~=~ ( \pa_{\rm K} \xi^{\rm I} ) \,
{\Bar \z}{}^{\rm K}_{\ad} ~~~.
\eqno(4.5)  $$
In other words, the undotted physical spinors and the dotted physical
spinors in a chiral-nonminmal (CNM) model are elements of ${}^* T_p({\cal 
M})$.  This situation can {\underline {never}} be realized with 
the sole use of chiral superfields! For purely chiral multiplet 
theories, the dotted spinors are elements of ${}^* {\Bar T}_p({\cal M})$.

The results of (4.4) indicate that the physical fields of the nonminimal
multiplet satisfy 
$$
( B, \, {\Bar \z}_{\Dot \a} \,) ~\in ~ {}^* T_p({\cal M}) ~~~,~~~
( {\Bar B}, \, {\z}_{\a} \,) ~\in ~ {}^* {\Bar T}_p({\cal M}) ~~~.
\eqno(4.6) $$
It can be seen that the geometrical interpretation here is totally
different from that of a chiral multiplet.  Further, we see that any
theory that includes {\underline {both}} chiral and 
nonminimal multiplets can be said to be {\underline {holomorphically}}
vector-like with respect to the intrinsic definition of space-time 
chirality associated with the non-linear $\s$-model by simply making 
sure that equal numbers and representations of $\psi_{\a}$ and 
${\Bar \z}_{\ad}$ spinors are present on ${}^* T_p({\cal M})$ and the 
conjugate spinors on ${}^* {\Bar T}_p({\cal M})$.

Let us say a few words about why we assume that the isometries are
generated by holomorphic co-tangents. As written (4.3) is a variation
for an infinitesimal transformation. To obtain the finite change
of variables associated with (4.3) requires exponentiation. In order
for this to be well defined, the co-tangent fields ($\xi_{(A)}^{\rm 
I}$) must form a ring.   This property would be violated if the 
co-tangent fields depended on both $\Phi$ and $\S$. It is a basic 
fact of 4D, N = 1 supersymmetry that the multiplication of chiral 
superfields forms a ring. This does not apply to the multiplication 
of nonminimal superfields.   Also in order for the finite transformations 
associated with the infinitesimal variations in (2.4) to form a group 
we cannot permit the co-tangent fields to depend on $\S$.

Now we apply the isometry variation to (2.3). That action will be
invariant if 
$$
\Big[~  \xi^{\rm L} \der{\Phi^{\rm L}} ~+~ {\Bar \xi}^{\Bar {\rm L}}
\der{{\Bar \Phi}^{{\Bar \rm L}}} ~+~ \S^{\rm K} ( \partder{\xi^{\rm 
I}}{\Phi^{\rm K}} )  \der{\S^{\rm I}} ~+~ {\Bar \S}^{\Bar {\rm K}}( 
\partder{{\Bar \xi}^{\Bar {\rm I}}}{{\Bar \Phi}^{\Bar {\rm K}}} ) 
\der{{\Bar \S}^{\Bar {\rm I}}} ~  \Big] {\Hat \O}~=~ \eta ~+~ {\Bar \eta}
 ~~~, \eqno(4.7) $$
for some chiral superfield $\eta$.

Of slightly more interest is the effect of the isometry variation on
the higher derivative terms. It can be seen that with a slight modification
of the definition of ${\Hat {\cal G}}^{ {\rm K}_j}_{{\underline c_1} ... 
{\underline c}_{k_j}}$ so as to respect the geometry of the K\" ahler
manifold (i.e. replace the derivatives by appropriate K\" ahler manifold
covariant derivatives\footnote{See discussion in chapter seven.}), the 
isometry variation of the Lagrangians takes the form
$$
\d_{\xi} {\cal L}_{(N)} ~\equiv~ P^{k_i \, {\underline c}_{1}  \, ...  
}_{A \, {\ad}_1 \, {\Dot \b}_1 \, ... } {\cal L}_{\xi} ( {\cal J}_{\rm I_1 
\, J_1 \, ... \,  K_1 \, ...}^{A \, k_i} (\Phi)) (  \prod 
{\cal G}^{ {\rm I}_i \,{\rm J}_i ~ {\ad}_i \, {\bd}_i} ) ( \prod
{\Hat {\cal G}}^{ {\rm K}_j}_{{\underline c_1} ... {\underline 
c}_{k_j}}) ~~~~,
\eqno(4.8)  $$
where ${\cal L}_{\xi}$ denotes the Lie derivative. Thus, requiring the
Lie derivative to vanish we find the higher derivative terms are
invariant under the isometries of the $\s$-model term.

The vanishing of the Lie derivative on ${\cal J}_{[ \rm {I \, J \, K \, 
L} ]}$, however, is too strong a condition. If we assume the weaker 
condition that its Lie derivative is equal to the exterior derivative 
of a holomorphic 3-form, then the WZNW term changes by a total divergence 
under the isometry variation.

\section{Super P-form Geometric Interpretation}

~~~~Some years ago \cite{PFM,pfmcon}, the complete simplex of 
{\underline {irreducible}} p-form gauge superfields was derived 
for 4D, N = 1 superspace.  We now wish to show that the CNM 
formulation of the WZNW action has a natural interpretation in 
the formalism of 4D, N = 1 super $p$-forms.  Since the component 
WZNW term is based on a 4-form, the natural starting place for 
our considerations is a super 4-form, $\O_{\un A \, \un B \, \un C 
\, \un D}^{WZNW}$ whose vector-vector-vector-vector component 
contains the usual component-level WZNW 4-form.   As noted in 
{\it {Superspace}} \cite{pfmcon}, in general for a super $p$-form, 
``(field strength) coefficients with more than two spinor indices 
have too low dimensions to contain component field strengths (or 
auxiliary fields), and must be constrained to vanish.''   We will 
call this the ``$p$-form theorem.\footnote{Interestingly enough, a 
slightly modified version of the $p$-form theorem holds for the 
case of 4D, \newline ${~~~\,~}$ N = 4 supergravity \cite{SG4} 
and 10D, N = 1 supergravity even with lowest order string corrections 
\newline ${~~~\,~}$ \cite{AAA,CCC}.}'' The simplex is illustrated
in the table below taken from {\it {Superspace}}.
\begin{center}
\renewcommand\arraystretch{1.2}
\begin{tabular}{|c|c|c| }\hline
${\rm p}$ & ${\Hat A}_p$ & ${\Hat {d A_p}}$
\\ \hline \hline
$ ~~0 ~~$ &  $ ~~ \Phi~~$ &  $ ~~i ( {\Bar \Phi}
- \Phi ) ~~$ \\ \hline
$ ~~1 ~~$ &  $ ~~ V ~~$ &  $ ~~i {\Bar D} {}^2 D_{\a} V ~~$ \\ \hline
$ ~~2 ~~$ &  $ ~~ \varphi_{\a}~~$ &  $ ~~
- \frac 12 (D^{\a} \varphi_{\a} + {\Bar D} {}^{\ad} {\Bar \varphi}_{\ad})
 ~~$ \\ \hline
$ ~~3 ~~$ &  $ ~~{\Hat V} ~~$ &  $ ~~{\Bar D}{}^2 {\Hat V} ~~$ \\ \hline
$ ~~4 ~~$ &  $ ~~ {\Hat \Phi}~~$ &  $ ~~ 0~~$ \\ \hline
\end{tabular}
\end{center}
\centerline{{\bf Table III}}
This table list for each value of $p$ the {\underline {irreducible}}
$p$-form and its irreducible super exterior derivative. One of the
most interesting features of this table is that it {\it {defines}} the
chiral superfield as the irreducible $0$-form.  The general real superfield
$V$ is associated with the the 1-form.

The plan of this chapter is to first recall all known irreducible
4D, N = 1 $p$-form theories. Next we use these irreducible $p$-forms
to study an assortment of super 4-forms that can be constructed
by using the super wedge product on the lower order forms. The main
purpose of this exercise is to establish a basis for the {\underline
{gauged}} CNW-WZNW term. Since the non-supersymmetric theory
of the gauged WZNW term is also a 4-form, it is natural to
expect that the supersymmetric theory should be related to
super 4-form geometry.

In the case of the 1-form gauge superfield, the field strength is 
a super 2-form $F_{\un A \, \un B }$. This corresponds to
the well studied case of supersymmetric Yang-Mills theory. As
such no few review comments are necessary in this case.
However, it is useful to note that given a constant algebraic
tensor ${\rm t}_{\rm {I \, J \, K \, L}}$ that satisfies some 
restrictions (see discussion above (6.1)), the superfield action
given by
$$
{\cal S}_{4-{\G}} ~=~  \int d^4 x d^2 \q \, d^2 
{\Bar \q} ~ {\rm t}_{\rm {I \, J \, K \, L}} ~ V^{\rm I}\,  \G^{\a \, 
\rm J} \G_{\un a}{}^{\rm K} \G^{\ad \, \rm L} ~~~,
\eqno(5.1) $$
will contain the component level term $\e^{\un a \, \un b \, \un c 
\, \un d} {\rm t}_{\rm {I \, J \, K \, L}} A_{\un a}{}^{\rm I} A_{\un 
b}{}^{\rm J} A_{\un c}{}^{\rm K}  A_{\un d}{}^{\rm L}$. This particular 
superfield action is distinguished by its very close relation with 
a term in the 3D, N = 1 superfield Chern-Simons action \cite{3dcs}.
(We have also written this term in the $K$-guage where all superconnections
$\G_{\un A}{}^{\rm I}$ are non-vanishing.)

The ordinary instanton number density is also associated with a component 
level 4-form. This suggests that there must be a corresponding super 
4-form associated with the instanton density of 4D, N = 1 supersymmetric 
Yang-Mills theory. So as an example of an irreducible super 4-form, 
it is interesting to treat this case.  We also recall that the 4D, 
N = 1 supersymmetric instanton number is also a chiral action and note 
that 4D, N = 1 supersymmetric Yang-Mills theory has one graded commutator
of the form
$$ [~ {\nabla}_{\a} \, , \, {\nabla}_{\un b}~ \} ~=~
i F_{\a \, \un b} ~=~ C_{\a \b} {\Bar W}_{\bd} ~~~. \eqno(5.2) $$
The field strength superfield $F_{\un A \, \un B}$ is an
irreducible super 2-form if the usual constraints of supersymmetric
Yang-Mills theory are enforced.  The super 2-form can be used to
construct a super 4-form ${\rm Y}_{\un A \, \un B \, \un C \, \un D}
\equiv F_{[ \un A \, \un B |}F_{| \un C \, \un D )}$. The only non-vanishing
components of this super 4-form are given by
$$ \eqalign{
{\rm Y}_{\a \, \b \, \un c \, \un d} &=~ - C_{\Dot \g \Dot \d}
C_{\a ( \g } C_{\d ) \b}  {\Bar W}^{\Dot \e} \, {\Bar W}_{\Dot \e}  ~~~, \cr
{\rm Y}_{\a \, \ad ~ \un c \, \un d} &=~ i 4 \e_{\un a \, 
\un c \, \un d  \un e \,} \, W^{\e} \, {\Bar W}^{\Dot \e}  ~~~, \cr
{\rm Y}_{\a \, \un b \, \un c \, \un d} &=~ - i 2
[ \, C_{\a \b} {\Bar W}_{\Dot \b} F_{\un c \, \un d} ~+~  
C_{\a \g} {\Bar W}_{\Dot \g} F_{\un d \, \un b} ~+~ 
C_{\a \d} {\Bar W}_{\Dot \d} F_{\un b \, \un c} ~\, ]    ~~~, \cr
{\rm Y}_{\un a \, \un b \, \un c \, \un d} &=~  \frac 1{4} 
F_{[ \un a \, \un b |}F_{| \un c \, \un d ] }  
~~~,} \eqno(5.3) $$
and their complex conjugates. 
Also in analogy with the super 4-form field strength, we here find $D_{\e} 
{\rm Y}_{\a \, \b \, \un c \, \un d} = 0$. The super 4-form $
{\rm Y}_{\un A \, \un B \, \un C \, \un D} $, like $F_{\un A \, \un B \, 
\un C \, \un D}$, is super closed and can be expressed as the
super exterior derivative of the super Chern-Simons 3-form
${\rm Y}_{\un A \, \un B \, \un C \, \un D} = D_{[\un A |}  X_{| \un B 
\, \un C \, \un D )}^{CS} ~-~ T_{[ \un A \, \un B| } {}^{\un F} X_{\un F | 
\, \un C \, \un D ) }^{CS}$ whose non-vanishing components are given by
$$
\eqalign{
X^{CS}_{\a \, \b \, \g} &\equiv ~ -  i {1 \over 3}  f_{\rm {I \, J \, K}}
        {\G}_{(\a|} {}^{{\rm I}} {\G}_{|\b|} {}^{{\rm J}}
         {\G}_{|\g) } {}^{{\rm K}}
 ~~~, \cr
X^{CS}_{\a \, \b \, \Dot \g} &\equiv ~  - i {1 \over 3} f_{\rm {I \, J \, K}}
        {\G}_{(\a|} {}^{{\rm I}} {\G}_{|\b|} {}^{{\rm J}}
         {\G}_{|\g) } {}^{{\rm K}}
 ~~~, \cr
X^{CS}_{\a \, \b \, \un c} &\equiv ~ {\G}_{(\a} {}^{{\rm I}} F^{\rm I}_{\b) 
       \un c } - i {1 \over 3}  f_{\rm {I \, J \, K}}
        {\G}_{(\a|} {}^{{\rm I}} {\G}_{|\b|} {}^{{\rm J}}
         {\G}_{|\g) } {}^{{\rm K}}
 ~~~, \cr
X^{CS}_{\a \, \bd \, \un c } &\equiv ~ {\G}_{\a} {}^{{\rm I}} F^{\rm I}_{\bd  
       \un c } + {\G}_{\bd} {}^{{\rm I}} F^{\rm I}_{\a  \un c }   
- i 2 f_{\rm {I \, J \, K}} {\G}_{\a} {}^{{\rm I}} 
        {\G}_{\bd} {}^{{\rm J}} {\G}_{\un c }  {}^{{\rm K}}
 ~~~, \cr
X^{CS}_{\a \, \un b \, \un c } &\equiv ~ {\G}_{\a } {}^{{\rm I}} 
        F^{\rm I}_{\un b \un c} - {\G}_{[ \un b |} {}^{{\rm I}} F^{\rm 
        I}_{\a | \un c ] }  -  i 2 f_{\rm {I \, J \, K}} {\G}_{\a }  
        {}^{{\rm I}} {\G}_{[ \un b|} {}^{{\rm J}} {\G}_{|\un c ]} 
        {}^{{\rm K}} 
        ~~~, \cr
X^{CS}_{\un a \, \un b \, \un c } &\equiv ~ \ha {\G}_{[\un a} {}^{{\rm I}} 
       F^{\rm I}_{\un b \un c] } - i {1 \over 3} f_{\rm {I \, J \, K}}
        {\G}_{[\un a|} {}^{{\rm I}} {\G}_{|\un b|} {}^{{\rm J}} {\G}_{|\un c] }
       {}^{{\rm K}}
~~~, }
\eqno(5.4) $$
where $\G_{\un A}^{~~ \rm I}$ denotes the superspace Yang-Mills
connection superfield. 

The instanton density action takes the super-geometric form 
$$\eqalign{
{\cal S}_{{Inst}} &=~ \Big[ ~- i \frac 1{12} \int d^4 x \, d^2 {\Bar 
\q} ~ \e^{\un a \, \un b \, \un c \, \un d} \,  C_{\ad \bd} \, 
{\rm Y}_{\a \, \b \, \un c \, \un d} ~+~ {\rm {h.\, c.}} ~ \,
 \Big]  ~~~,  \cr
&=~ \Big[ ~ i \frac 1{4} \int d^4 x \, d^2 {\Bar \q} ~ \e^{\un a \, 
\un b \, \un c \, \un d} \,  C_{\ad \bd} \, F_{\a \, \un c} \, 
F_{\b \, \un d} ~+~ {\rm {h.\, c.}} ~ \,  \Big] ~~~, \cr
&=~ \Big[ ~   i \frac 1{4}
\int d^4 x \, d^2 {\Bar \q} ~ C^{\Dot \g \Dot \d} C^{\a \g} C^{\b \d}
F_{\a \, \un c} \, F_{\b \, \un d} ~+~ {\rm {h.\, c.}} ~ \,
 \Big]  ~~~, \cr
&=~ \Big[ \, i \int d^4 x \, d^2 {\q} ~  W^{\a} W_{\a}  ~+~ {\rm {h.
\, c.}}\, ~ \Big]  ~~~, } \eqno(5.5) $$
and we see that there is an exact analogy between the
forms of the corresponding non-supersymmetric quantities and
their supersymmetric extensions!

In the case of the 2-form gauge superfield, the field strength is 
a super 3-form $G_{\un A \, \un B \, \un C }$ that is known to satisfy 
the following constraints,
$$
G_{\a \, \b \, \un C} ~=~ 0 ~~~,~~~
G_{\a \, \bd \, \un c} ~=~ i C_{\a \g} C_{\Dot \b  \Dot \g} G ~~~, 
\eqno(5.6) $$
where $D^2 G = {\Bar D}^2 G = G - {\Bar G} = 0$.  The remaining non-vanishing
field strengths can be written as,
$$ \eqalign{
G_{\a \, \un b \, \un c } &=~ C_{\a  (\b} C_{\g )  \d} C_{\Dot \b  \Dot \g}
{D}^{\d} G ~~~, \cr
G_{\un a \, \un b \, \un c } &=~ \e_{\un a \, \un b \, \un c \, \un d} 
\, [ \, {D}^{\d} \, , \, {\Bar D}{}^{\Dot \d}  ~ ] G
~~~. }\eqno(5.7)  $$
In this last expression we have utilized the definition
$$\e_{\un a \, \un b \, \un c \, \un d} \equiv~  i \,  \frac 12 ~ 
[ ~ C_{\a \b} C_{\g \d} C_{ \Dot \a (\Dot \g } C_{\Dot \d ) \Dot \b}
- C_{\Dot \a \Dot \b} C_{\Dot \g \Dot \d} C_{ \a ( \g } C_{ \d )  
\b} ~] ~~~. 
\eqno(5.8) $$
We also know that the field strength can be expressed in terms of
an unconstrained chiral spinor superfield $\varphi^{\a}$ (analogous
to $V$ for Yang-MIlls theory) via $G = - \frac 12 (\, D^{\a} \varphi_{\a}
+ {\Bar D}{}^{\ad} {\Bar \varphi}_{\ad}\,)$.  We note in passing that this is 
the rigid super-geometrical formulation of the axion multiplet (also 
known as the linear multiplet) that contains the axion, dilaton and 
dilatino.   The action for this multiplet that contains purely the square 
of the 3-form field strength is just 
$$
{\cal S}_{2-{\rm form}} ~=~ - \frac 12 \int d^4 x d^2 \q \, d^2 {\Bar \q} ~ 
G^2  ~~~~.
\eqno(5.9) $$
In the presence of a vector multiplet, an interesting interaction of
the form
$$
{\cal S}_{2-{\rm form}-{\rm {mass}}} ~=~ \int d^4 x d^2 \q  ~ 
V G \, + \, {\rm {h.\,c.}} ~=~ - \int d^4 x d^2 \q  ~ \varphi^{\a} W_{\a} 
\, + \, {\rm {h.\,c.}}~~~,
\eqno(5.10) $$
which contains the bosonic term $\e^{\un a \, \un b \, \un c \, \un d}
A_{\un a} \pa_{\un b} b_{\un c \un d}$ (where $A_{\un a}$ is the
component vector in the vector multiplet and $b_{\un c \un d}$ is the
component tensor in the 2-form multiplet) can be introduced.

In the case of the 3-form gauge superfield, the field
strength is a super 4-form $F_{\un A \, \un B \, \un C \, \un D}$ that is 
known to satisfy the following
constraints,
$$
F_{\a \, \b \, \g ~ \un D} ~=~ F_{\ad ~ \b \, \g \, \un D} ~=~
F_{\ad \, \b \, \un c \, \un d} ~=~ 0 ~~~,~~~
F_{\a \, \b \, \un c \, \un d} ~=~  C_{\Dot \g \Dot \d} C_{\a (\g}
C_{\d ) \b} {\Bar {\cal F}} ~~~, 
\eqno(5.11) $$
where ${\Bar D}_{\ad} {\cal F} = 0$. The remaining non-vanishing
field strength superfields take the forms
$$ \eqalign{
F_{\a \, \un b \, \un c \, \un d} &=~ - 
\e_{\un a \, \un b \, \un c \, \un d} {\Bar D}^{\ad} {\Bar {\cal F}} 
 ~~~, \cr
F_{\un a \, \un b \, \un c \, \un d} &=~ i 
\e_{\un a \, \un b \, \un c \, \un d} \Big[ \, {D}^2 {\cal F}
~-~ {\Bar D}^2 {\Bar {\cal F}} ~ \Big]
~~~. }\eqno(5.12)  $$
As was described in the first considerations of irreducible super 
$p$-forms \cite{PFM}, the super 4-form defined by (5.11) and (5.12) is 
super-closed $((d {F})_{\un A \, \un B \, \un C \, \un D 
\, \un E} = 0$).   The action for this multiplet that contains 
the square of the 4-form field strength is just 
$$
{\cal S}_{4-{\rm form}} ~=~  \int d^4 x d^2 \q \, d^2 {\Bar \q} ~ 
{\Bar {\cal F}} \, {\cal F}
~~~~,~~~
\eqno(5.13) $$
and as well a purely topological term is obtained from
$$
{\cal S}_{3-{\rm form}-{\rm top}} ~=~ i \,  \int d^4 x d^2 \q  ~ 
 {\cal F} ~+ {\rm {h.\,c.}}  ~~~.
\eqno(5.14) $$
The field strength ${\cal F}$ can be expressed as ${\cal F}
= {\Bar D}{}^2 {\Hat V}$ in terms of an unconstrained prepotential
superfield ${\Hat V}$.

In the case of the 4-form gauge superfield, the field strength 
is a super 5-form ${\cal H}_{\un A \, \un B \, \un C \, \un D \, 
\un E}$ which we may constrain to be identically zero.  In this
case, there exist a super closed 4-form gauge superfield denoted 
by ${\rm h}_{\un A \, \un B  \, \un C \, \un D}$ that is not 
super exact.  This 4-form gauge superfield takes the form,
$$ \eqalign{
{\rm h}_{\a \, \b \, \g ~ \un D} &=~ {\rm h}_{\ad ~ \b \, \g \, \un D} ~=~
{\rm h}_{\ad \, \b \, \un c \, \un d} ~=~ 0 ~~~,~~~ \cr
{\rm h}_{\a \, \b \, \un c \, \un d} &=~  C_{\Dot \g \Dot \d} C_{\a (\g}
C_{\d ) \b} {\Hat {\Bar {\Phi}}} ~~~, \cr
{\rm h}_{\a \, \un b \, \un c \, \un d} &=~ - 
\e_{\un a \, \un b \, \un c \, \un d} {\Bar D}^{\ad} {\Hat {\Bar {\Phi}}} 
 ~~~, \cr
{\rm h}_{\un a \, \un b \, \un c \, \un d} &=~ i 
\e_{\un a \, \un b \, \un c \, \un d} \Big[ \, {D}^2 {\Hat {\Phi}}
~-~ {\Bar D}^2 {\Hat {\Bar {\Phi}}} ~ \Big]
~~~. }\eqno(5.15)  $$
where ${\Bar D}_{\ad} {\Hat {\Phi}} = 0$.

In closing this section of this chapter, we emphasize that the 
super 4-forms discussed above bare no direct relation to the super 
WZNW-form discussed next. The preceding discussion was included as 
a convenient introduction to the topic of {\underline {irreducible}} 
4D, N = 1 super 4-forms.   The most important lesson to draw from 
this discussion is that whenever ordinary $p$-forms appear in a 
non-supersymmetric theory, it should be expected that irreducible 
4D, N = 1 super $p$-forms will appear in the supersymmetric extension 
of that theory.

The ``$p$-form theorem'' together with the result above suggests that 
the first non-trivial component relevant to the irreducible WZNW super 
4-form is the component $\O_{\a \, \b \, \un c \, \un d}^{WZNW}$.  
All lower or equal dimension components must be equal to zero up to an 
exact super 4-form.  We can thus make the natural ansatz
$$
\O_{\a \, \b \, \un c \, \un d}^{WZNW} ~=~ \frac1{4!} {\Bar \O}_{[\rm{I 
\, J \, K \, L}\,]} (D_{\a} {\Bar \S}^{\rm I} ) (D_{\b} {\Bar \S}^{
\rm J} ) (\pa_{\un c} {\Bar \Phi}^{\rm K}) (\pa_{\un d} {\Bar \Phi}^{\rm 
L}) ~~~,  \eqno(5.16)  $$
which has all the correct symmetries to be identified as the first
non-trivial component of WZNW super 4-form.  If the coefficient
function ${\Bar \O}_{[\rm{I \, J \, K \, L}\,]}$ is anti-holomorphic,
then we find, $D_{\e} \O_{\a \, \b \, \un c \, \un d}^{WZNW} = 0$
exactly like the two cases discussed earlier in the chapter.
Clearly this 4D, N = 1 super 4-form component has the obvious 
interpretation as being the {\underline {anti}}-{\underline {holomorphic}} 
pull-back of the anti-holomorphic K\" ahler manifold tensor ${\Bar \O}_{
[\rm{I \, J \, K \, L}\,]}$.  It can be noted at this stage that if we were
to restrict ourselves solely to chiral superfields, the only anti-holomorphic
pull-back would have four vector indices that could be contracted
with $\e_{\un a \, \un b \, \un c \, \un d}$. The resultant superfield
could then be used in an action. However, the component level WZNW term
is absent after integrating out the Grassmann coordinates.

With a bit of algebra, we find that the general irreducible decomposition 
with respect to Lorentz symmetry yields,
$$
\O_{\a \, \b \, \un c \, \un d}^{WZNW} ~\equiv~ {\Hat \O}_{( \a \, \b \, 
\g \, \d )}^{(W)}  C_{\Dot \g \, \Dot \d} ~+~ {\Hat \O}_{( \a \, \b ) \, 
(\Dot \g \, \Dot \d)}^{(R)}  C_{\g \, \d} ~+~ {\Hat \O}^{(S)} C_{\g \, 
( \a} C_{\b ) \, \d} C_{\Dot \g \, \Dot \d}
~~~,
\eqno(5.17)  $$
where the labels $(W)$, $(R)$ and $(S)$ are meant to be reminders of
the similarity of this decomposition to that of the Riemann curvature
tensor into its Weyl, Ricci (traceless) and scalar curvature
components. To complete our proposed super 4-form description, we
require explicit expressions for $\O_{\a \, \un b \, \un c \, \un d
}^{WZNW}$ and $\O_{\un a \, \un b \, \un c \, \un d}^{WZNW}$.  For
these we propose,
$$
\O_{\a \, \un b \, \un c \, \un d}^{WZNW} ~=~ \e_{\un b 
\, \un c \, \un d}{}^{\, \un e} [~ C_{\a \e} {\Bar D}_{\Dot \e} 
{\Hat \O}^{(S)} ~-~ \frac 13 {\Bar D}^{\Dot \l}
{\Hat \O}_{( \a \, \e ) \, (\Dot \l \, \Dot \e)}^{(R)} ~]
~\equiv~  \e_{\un b \, \un c \, \un d}{}^{\, \un e} \, 
{\Hat \O}_{\un e \, \a} ~~~,
\eqno(5.18)  $$
$$
\O_{\un a \, \un b \, \un c \, \un d}^{WZNW} ~=~ 
i  \e_{\un a \, \un b \, \un c \, \un d} \Big[ \, {D}^2 
{\Hat {\Bar \O}}{}^{(S)} ~-~ {\Bar D}^2 {\Hat \O}^{(S)} \, \Big]
~~~. {~~~~~~~~} {~~~~~~~~} {~~~~~~~~~~~~~} 
\eqno(5.19)  $$
Thus (5.16-5.19) complete our ansatz for the super 4-form 
$\O_{\un A \, \un B \, \un C \, \un D}^{WZNW}$ and with it we can easily 
compute its super exterior derivative via the equation,
$$
(d \O^{WZNW})_{\un A \, \un B \, \un C \, \un D \, \un E} ~=~ 
D_{[\un A |}  \O_{| \un B \, \un C \, \un D\, \un E )}^{WZNW}
~-~ T_{[ \un A \, \un B| } {}^{\un F} \O_{\un F | \, \un C \, 
\un D\, \un E ) }^{WZNW} ~~~,
\eqno(5.20) $$
which when expressed in terms of the non-vanishing components
of $ \O_{\un A \, \un B \, \un C \, \un D}^{WZNW}$ takes
the more explicit forms,
$$ \eqalign{ {~~~~~~~~} 
(d \O^{WZNW})_{\ad \, \a \, \g \, \un d \, \un e} &\equiv ~ 
       {\Bar D}_{\ad} \, \O_{\a \, \g \, \un d \, \un e}^{WZNW} 
        ~+~ i\,  \O_{(\g \, \a) \ad \,  \un d \, \un e}^{WZNW} ~~~, \cr
(d \O^{WZNW})_{ \a \, \b \, \un c \, \un d \, \un e } &\equiv ~ 
        \frac 12 \pa_{[ \un c |}  \O_{ \a \, \b \, | \un d \, \un e]}^{WZNW}
        ~+~ D_{(\a} \O_{\b ) \un c \, \un d \, \un e}^{WZNW} ~~~, \cr
(d \O^{WZNW})_{\ad \, \b \, \un c \,  \un d \, \un e } &\equiv ~ 
        {\Bar D}_{\ad } \O_{\b \, \un c \, \un d \, \un e}^{WZNW}
         ~+~ D_{\b} \O_{\ad \, \un c \, \un d \, \un e}^{WZNW}
         ~-~ i \, \O_{\b \ad \, \un c \, \un d \, \un e}^{WZNW} ~~~, \cr
(d \O^{WZNW})_{\a \, \un b \, \un c \, \un d \, \un e} &\equiv ~ 
         D_{\a} \O_{ \un b \, \un c \, \un d \, \un e}^{WZNW}
         ~-~ \frac 16 \pa_{[ \un b |} \O_{\a | \un c \, \un d \, 
         \un e]}^{WZNW} ~~~, \cr
(d \O^{WZNW})_{\un a \, \un b \, \un c \, \un d \, \un e} &\equiv ~ 
\frac1{4!} \pa_{[ \un a |} \O_{ | \un b \, \un c \, \un d \, 
\un e]}^{WZNW}
~~~. }
\eqno(5.21) $$
These yield after substitution from (5.10-5.13) the results
$$ \eqalign{ {~~} 
(d \O^{WZNW})_{\ad \, \a \, \g \, \un d \, \un e} &\equiv ~ 
         C_{\Dot \d \Dot \e} {\Bar D}_{\Dot \a} {\Hat \O}_{( \a \, 
         \g \, \d \, \e )}^{(W)} \,
         + \, \frac 13 C_{\d \e} [\, {\Bar D}_{\Dot \a} {\Hat 
         \O}_{( \a \, \g ) \, (\Dot \d \, \Dot \e)}^{(R)} \, + \, 
         {\Bar D}_{\Dot \e} {\Hat \O}_{( \a \, \g ) \, 
         (\Dot \a \, \Dot \d)}^{(R)} \cr 
          &{~~~~} + \, {\Bar D}_{\Dot \d} 
         {\Hat \O}_{( \a \, \g ) \, (\Dot \a \, \Dot \e)}^{(R)} \,  \, ] \,
         + \, \frac 16  C_{\Dot \d \Dot \e}  [\, C_{\e \a}
         {\Bar D}{}^{\Dot \l} {\Hat \O}_{( \a \, \g ) \, (\Dot \l \, \Dot \a)}^{(R)}
         \, + \, C_{\e \g} {\Bar D}{}^{\Dot \l} {\Hat \O}_{( \a \, \d ) \, (\Dot 
         \l \, \Dot \a)}^{(R)} \cr 
         &{~~~~} \, +  \, C_{\d \a} {\Bar D}{}^{\Dot \l} {\Hat 
         \O}_{( \g \, \e ) \, (\Dot \l \, \Dot \a)}^{(R)} \, + \, C_{\d \g}  
         {\Bar D}{}^{\Dot \l} {\Hat \O}_{( \a \, \d ) \, (\Dot \l \, \Dot \a)}^{(R)}
         \, ] ~~~, \cr 
(d \O^{WZNW})_{ \a \, \b \, \un c \, \un d \, \un e } &\equiv ~ -
          \e_{\un c \, \un d \un e}{}^{\un k} \, [ ~ \pa^{\l} {}_{\Dot \k}
          {\Hat \O}_{( \a \, \b \, \k \, \l )}^{(W)} \, - \, 
           \pa_{\k} {}^{\Dot \l} {\Hat \O}_{( \a \, \b ) \, (\Dot 
           \k \, \Dot \l)}^{(R)} \cr
           &{~~~~} \, +  \, \frac 13 ( \, \pa_{\a} {}^{\Dot \l} 
           {\Hat \O}_{( \b \, \k ) \, (\Dot \k \, \Dot \l)}^{(R)} \, + \, 
           \pa_{\b} {}^{\Dot \l} {\Hat \O}_{( \a \, \k ) \, (\Dot 
           \k \, \Dot \l)}^{(R)} \, ) ~ ] 
           ~~~, \cr
(d \O^{WZNW})_{\ad \, \b \, \un c \,  \un d \, \un e } &\equiv ~ 
           \frac 16 \e_{\un c \, \un d \un e}{}^{\un k}\, [~ {\Bar D}{}^{\Dot \l}
           {\Bar D}_{\Dot \l} {\Hat \O}_{( \b \, \k ) \, (\Dot \a \, \Dot \k)}^{(R)}
           \, + \, {\rm {h.\, c.}} ~] 
           ~~~, \cr
(d \O^{WZNW})_{\a \, \un b \, \un c \, \un d \, \un e} &\equiv ~
          - \frac 13 \pa^{\un k} \, {\Bar D}{}^{\Dot \l}  {\Hat \O}_{( \a \, \k ) 
          \, (\Dot \k \, \Dot \l)}^{(R)}
          ~~~, \cr
(d \O^{WZNW})_{\un a \, \un b \, \un c \, \un d \, \un e} &\equiv ~ 
         i \frac 1{4!} \e_{[ \un a \, \un b \, \un c \, \un d |} \,
         \pa_{| \un e ] } \Big[ \, {D}^2 {\Hat {\Bar \O}}{}^{(S)} ~-~ {\Bar 
         D}^2 {\Hat \O}^{(S)} \, \Big]
~~~. }
\eqno(5.22) $$
The differences between the closed super $p$-forms described by
$F_{\un A \, \un B \, \un C \, \un D}$ or ${\rm Y}_{\un A \, \un B \, 
\un C \, \un D}$ and the non-closed super $p$-form described by
$\O^{WZNW}_{\un A \, \un B \, \un C \, \un D}$ are now completely 
obvious. The super exterior derivative of the former vanish while
the super exterior derivative of the latter depends only on its
`Weyl' and `traceless Ricci' pieces. For the former $(d F)_{\un a 
\, \un b \, \un c \, \un d \, \un e} = (d {\rm Y})_{\un a \, 
\un b \, \un c \, \un d \, \un e} = 0$, while $(d \O^{WZNW})_{\un a 
\, \un b \, \un c \, \un d \, \un e} \neq 0$.

A final point to note is that in the CNM formulation, the 4D, N = 1 
supersymmetric WZNW term can be expressed as
$$ \eqalign{ {~~~~~~~~}
{\cal S}_{{\rm H.\, D.}}^{WZNW} 
&=~ \Big[ ~ - \frac 1{12} \int d^4 x \, d^2 {\Bar \q} ~ \e^{\un a \, 
\un b \, \un c \, \un d} \, C_{\ad \bd} \, \O_{\a \, \b \, \un c \, 
\un d}^{WZNW} ~+~ {\rm {h.\, c.}} ~ \,  \Big] ~~~, \cr
&=~  \Big[ ~- \frac 1{12}
\int d^4 x \, d^2 {\Bar \q} ~ C^{\Dot \g \Dot \d} C^{\a \g}
C^{\b \d} \O_{\a \, \b \, \un c \, \un d}^{WZNW} ~+~ {\rm {h.\, c.}} ~
\, \Big]  ~~~, \cr
&=~ \Big[ ~\int d^4 x \, d^2 {\q} ~ {\Hat {\Bar \O}}{}^{(S)} ~+~ 
{\rm {h.\, c.}} ~ \,  \Big]
~~~, } \eqno(5.23) $$
so that our proposal is the chiral superspace integral of a super 4-form!
No such super-geometric interpretation is available for (3.7). Roughly
speaking, the result in (5.5) is analogous to that in (5.17) although
in the latter case the action is {\underline {not}} a surface term.
Alternately we see that there is clearly a close super geometrical
relation between the instanton density and the CNM-WZNW term.

\section{A Component Preview of 4D, N = 1 Super-\newline 
${}$symmetric  Auxiliary-free 4-${\cal J}$ Terms}

~~~~~The task of giving the complete component level description 
of the action in (2.3) is enormous. A major step is simply the
component evaluation of ${\cal S}_{\s}$. This is important in
order to be able to derive the auxiliary field equations of
motion among other reasons. This will be completed in a future
work. In this section, we wish to focus some attention on the
component structure of the lowest order terms in ${\cal S}_{\rm 
{H. D.}}$. From equation (2.7) we see that there are in fact
three broad classes of such terms. One class consists of terms
quadratic in ${\Hat {\cal G}}$. Such terms are purely proportional
to fermions. Another class of terms are linear in ${\Hat {\cal G}}$
and linear in ${\cal G}$. Finally, there terms quadratic in ${\cal G}$.
(an example of such a term can be seen in (A.8)).  It is only the 
second class for which we will give a brief description. This class 
of terms includes the Skyrme and WZNW terms. 

We begin with a fourth order holomorphic tensor denoted  by ${\cal 
J}_{\rm {I \, J \, K \, L}} (\Phi)$ which most generally satisfies ${\cal 
J}_{\rm {I \, J \, K \, L}} = {\cal J}_{\rm {J \, I \, L \, K}}$.  
In the applications we wish to show, we satisfy this condition
by imposing ${\cal J}_{\rm {I \, J \, K \, L}} = - {\cal J}_{\rm {J \, 
I \, K \, L}}$ and ${\cal J}_{\rm {I \, J \, K \, L}} = - {\cal J}_{\rm 
{I \, J \, L \, K}}$. The lowest order term we wish to
show thus arise as,
$$
{\cal S}_{{\rm H.\, D.}}^{(4)} ~=~   \,  \int d^4 x \, d^2 \q ~ {\cal 
J}_{\rm I \, J \, K \, L} (\Phi) ({\Bar D}^{\ad} \S^{\rm I} \, ) \, (
{\Bar D}^{ \bd} \S^{\rm J} \, ) \, (\pa^{\g} {}_{\ad} \Phi^{\rm K} \,) \, 
( \pa_{\g \bd} \Phi^{\rm L} \,)  ~+~ {\rm h.}{\rm c.} ~~~~.
\eqno(6.1) $$
However, (6.1) has the consequence that it describes more than the WZNW 
term.  As well it contains the Skyrme term. This occurs by noting that an 
irreducible decomposition of ${\cal J}_{\rm I \, J \, K \, L}$ takes the 
form
$${\cal J}_{\rm I \, J \, K \, L} ~=~ \sum_{{\rm A}} {\cal J}^{A}_{\rm I 
\, J \, K \, L } ~~~,
\eqno(6.2) $$
where ${\rm A}$ denotes the different irreducible representations.
(See also A.12.)

The calculation of the component results follows using the by now
well established projection technique. We find ${\cal S}_{{\rm H.\, 
D.}}^{(4)}$ leads to
$$ \eqalign{
 &   \int d^4 x d^2 \q ~ {\cal J}_{\rm I \, J \, K \, L} (
\Phi) ({\Bar D}^{\ad} \S^{\rm I} \, ) ({\Bar D}^{\Dot \b} \S^{\rm J} \, 
) (\pa^{\g}{}_{\ad} \Phi^{\rm K} \,) ( \pa_{\g \Dot \b} \Phi^{\rm L} \,) 
{~~~~~~~~~~}  \cr
 &=~     \int d^4 x \Big[ {~~}  {\cal J}_{\rm I \, J \, K \, L}  
\, ( \pa^{\a \ad} B^{\rm I} ~+~ i \,p^{\a \ad ~ \rm I} \, ) (  \pa_{\a}
{}^{\Dot \b} B^{\rm J} ~+~ i \, p_{\a} {}^{\Dot \b ~ \rm J} \, )
(\pa^{\g}{}_{\ad} A^{\rm K} \,) ( \pa_{\g \Dot \b} A^{\rm L} \,) \cr
& {~~~~~~~~~~~~~~~} +~   {\cal J}_{\rm I \, J \, K \, L} \, ( i \pa_{\a}
{}^{\ad} \r^{\a \, {\rm I}} ~-~ 2 \b^{\ad \, {\rm I}} \,) 
\,{\Bar \zeta}^{\Dot \b \, {\rm J}} (\pa^{\g}{}_{\ad} A^{\rm K} \,)
 ( \pa_{\g \Dot \b} A^{\rm L} \,) \cr
& {~~~~~~~~~~~~~~~} +~  {\cal J}_{\rm I \, J \, K \, L}  \,{\Bar \zeta}^{\Dot 
\a \, {\rm I}}{\Bar \zeta}^{\Dot \b \, {\rm J}} [\, (\pa^{\g}{}_{\ad} \psi^{\a 
\rm K} \,) ( \pa_{\g \Dot \b} \psi_{\a}{}^{\rm L} \,) ~+~ 2
(\pa^{\g}{}_{\ad} A^{\rm K} \,) ( \pa_{\g \Dot \b} F^{\rm L} 
\,) \,] \cr 
& {~~~~~~~~~~~~~~~} +~ i \, 4 {\cal J}_{\rm I \, J \, K \, L}  \,
(   \pa^{\a \ad} B^{\rm I} ~+~ i p^{\a \ad ~ \rm I} \, ) \, 
{\Bar \zeta}^{\Dot \b \, {\rm J}} (\pa^{\g}{}_{\ad} \psi_{\a}^{ 
\rm K} \,)  ( \pa_{\g \Dot \b} A^{\rm L} \,) \cr
& {~~~~~~~~~~~~~~~} +~  2 {\cal J}_{\rm I \, J \, K \, L \, , {\rm M}}  \,
\psi^{\a \rm M} \, {\Bar \zeta}^{\Dot  \a \, {\rm I}}{\Bar \zeta}^{\Dot \b \, 
{\rm J}} (\pa^{\g}{}_{\ad} \psi_{\a}^{\rm K} \,)  ( \pa_{\g \Dot \b} 
A^{\rm L} \,) \cr
& {~~~~~~~~~~~~~~~} -~ i 2 {\cal J}_{\rm I \, J \, K \, L \, , {\rm M}}  \,
\psi_{\a}{}^{\rm M} (   \pa^{\a \ad} B^{\rm I} ~+~ i p^{\a \ad ~ \rm I} 
\, ) \, {\Bar \zeta}^{\Dot \b \, {\rm J}} (\pa^{\g}{}_{\ad} A^{\rm K} \,) 
( \pa_{\g \Dot \b} A^{\rm L} \,) \cr
& {~~~~~~~~~~~~~~~} +~ {\cal J}_{\rm I \, J \, K \, L \, , {\rm M}}  \,
F^{\rm M} \,{\Bar \zeta}^{\Dot  \a \, {\rm I}}{\Bar \zeta}^{\Dot \b \, 
{\rm J}} (\pa^{\g}{}_{\ad} A^{\rm K} \,) ( \pa_{\g \Dot \b} A^{\rm L} 
\,)  \cr
& {~~~~~~~~~~~~~~~} +~ \frac 12 \, {\cal J}_{\rm I \, J \, K \, L \, , 
{\rm {M \, N}}}  \, \psi^{\a \, \rm M} \, \psi_{\a \,}{}^{\rm N} \,
{\Bar \zeta}^{\Dot  \a \, {\rm I}}{\Bar \zeta}^{\Dot \b \, 
{\rm J}} (\pa^{\g}{}_{\ad} A^{\rm K} \,) ( \pa_{\g \Dot \b} A^{\rm L} 
\,)  ~ \Big] 
~~~~. }
\eqno(6.3) $$
As can be seen, only the first line of the rhs consists of purely bosonic
terms.  Let us focus our analysis by only considering these terms.

It is our first observation that if we set the auxiliary field $p_{
\underline a}{}^{\rm I}$ to zero, then the purely bosonic terms collapse to
$$ \eqalign{   
\int d^4 x \, d^2 \q ~ &{\cal J}_{\rm I \, J \, K \, L} (\Phi) \,
({\Bar D}^{\ad} \S^{\rm I} \, ) \, ({\Bar D}^{\Dot \b} \S^{\rm J} \, ) \,
(\pa^{\g} {}_{\ad} \Phi^{\rm K} \,) \, ( \pa_{\g \Dot \b} \Phi^{\rm L} 
\,) |_{phys.\, fields}{~~~~~~~~~~}  \cr
 &=~  \int d^4 x \Big[ \,  {\cal J}_{\rm I \, J \, K \, L} (A) \, ( 
\pa^{\a \ad} B^{\rm I} \, ) \, ( \pa_{\a}{}^{\Dot \b} B^{\rm J} \, ) \,
(\pa^{\g}{}_{\ad} A^{\rm K} \,) \, ( \pa_{\g \Dot \b} A^{\rm L} \,) \,
\Big]  \cr
 &=~  \int d^4 x \Big[ \,  {\cal J}_{\rm I \, J \, K \, L} (A) \,
{\rm P}^{\un a \, \un b \, \un c \, \un d} \,
( \pa_{\un a} B^{\rm I} \, ) \, ( \pa_{\un b} B^{\rm J} \, )
\, (\pa_{\un c} A^{\rm K} \,) \, ( \pa_{\un d} A^{\rm L} \,) 
\, \Big]  ~~~~, }
\eqno(6.4) $$
where ${\rm P}^{\un a \, \un b \, \un c \, \un d}
\equiv \frac 12 [ \eta^{\un a \, \un b} \eta^{\un c \,
\un d} ~+~ \eta^{\un a [ \un c}  \eta^{\un d ] 
\un b} ~+~ i \e^{\un a \, \un b \, \un c  \,
\un d} ] $.  In going from the second to the third line above,
we have made use of the following identities,
$$\eqalign{
\eta_{\un a \un b} &\equiv~ C_{\a \b} C_{\ad \bd} ~~~,~~~
\eta^{\un a \un b} ~\equiv~ C^{\a \b} C^{\ad \bd} ~~~, \cr
C^{\a \b} C^{\g \d} C^{\ad \Dot \g} C^{\bd \Dot \d} &=~ \frac 12
 [~ \eta^{\un a \un b} \eta^{\un c \un d} ~+~ 
\eta^{\un a \un c} \eta^{\un d \un b} ~-~ 
\eta^{\un a \un d} \eta^{\un c \un b} ~+~ i\, \e^{\un a \, \un b \,
\un c \, \un d} ~] ~~~, }
\eqno(6.5) $$
Due to the symmetry restrictions on ${\cal J}_{\rm I \, J \, K \, L}$
imposed above equation (6.1) we may drop the leading term in the
definition of ${\rm P}^{\un a \, \un b \, \un c \, \un d}$ since
upon contraction this gives zero. The next term in ${\rm P}^{\un a 
\, \un b \, \un c \, \un d}$ produces the Skyrme term and the
remaining term produces the WZNW term (see also A.12 in appendix A).

Although we defer to the future a detailed discussion of the component
results implied by the CNM approach to the 4D, N = 1 supersymmetric
low-energy QCD effective action, there is one aspect that is so amusing
that we wish to focus upon it here.  The dynamical bosonic fields in our
construction correspond to the leading components of $\Phi^{\rm I}$ and
$\S^{\rm I}$.  For these we may write,
$$ \eqalign{ {~~~~~~~~~}
\Phi^{\rm I} | &=~ {A}^{\rm I}(x) ~=~ {\cal A}^{\rm I} (x)~+~ i 
\Big[ \, \Pi^{\rm I} (x) cos(\g_{\rm S} ) ~+~ \Theta^{\rm I} (x)
sin(\g_{\rm S} ) \, \Big] ~~~, \cr
\S^{\rm I} |  &=~ {B}^{\rm I}(x) ~=~ {\cal B}^{\rm I}(x) ~+~ i 
\Big[ \,  - \Pi^{\rm I} (x) sin(\g_{\rm S} ) ~+~ \Theta^{\rm I} 
(x) cos(\g_{\rm S} ) \, \Big] ~~~,}
\eqno(6.6) $$ 
in terms of two real octets of scalar spin-0 fields ${\cal A}^{\rm I}$
and ${\cal B}^{\rm I}$ as well as two real octets of psuedo-scalar spin-0 
fields ${\Pi}^{\rm I}$ and ${\Theta}^{\rm I}$.  Here we introduce a mixing
angle $\g_{\rm S}$ that is restricted by the form of the supersymmetric WZNW 
action to satisfy the condition $sin(2 \g_{\rm S} ) \neq 0$\footnote{
In our work of ref. \cite{SG} we simply assumed maximal mixing
for the sake of simplicity.} (see below).

The amusement begins by recalling the structure of the 
Glashow-Salam-Weinberg model \cite{GSW}.  We also can define an octet
of complex scalar spin-0 fields by the definitions,
$$ {G}^{\rm I}_+ ~\equiv~ \frac1{\sqrt 2}  [ ~ {\cal A}^{\rm I} ~+~ i 
{\cal B}^{\rm I} ~]  ~~~,~~~ {G}^{\rm I}_- ~\equiv~ \frac 1{\sqrt 2}  
[ ~ {\cal A}^{\rm I} ~-~ i {\cal B}^{\rm I} ~] ~~~,
\eqno(6.7) $$
so that ${G}^{\rm I}_- ~=~ [ {G}^{\rm I}_+ ]^*$. Finally we see that the
set of spin-0 fields $(\, \Pi^{\rm I}, \, \Theta^{\rm I} ,\, {G}^{\rm I}_+ 
,\, {G}^{\rm I}_- \, )$ bares an {\underline {uncanny}} resemblance to $(
A_{\m} ,\, Z^0_{\mu}, \, W^+_{\m} ,\,  W^-_{\m} \, )$ of the GSW model!  The 
mirth even more increases when we realize that after supersymmetry-breaking
there must develop a mass gap between $\Pi^{\rm I}$ and the remaining
spin-0 fields. Furthermore, the mixing angle here ($\g_{\rm S}$) is clearly
the analog of the weak mixing angle ($\q_{\rm W}$). Substituting (6.6)
into (6.4) and keeping only the purely  $\Pi$ dependent terms yields,
$$ \eqalign{   
\int d^4 x  &d^2 \q ~ {\cal J}_{\rm I \, J \, K \, L} (\Phi) \,
({\Bar D}^{\ad} \S^{\rm I} \, ) \, ({\Bar D}^{\Dot \b} \S^{\rm J} \, ) \,
(\pa^{\g} {}_{\ad} \Phi^{\rm K} \,) \, ( \pa_{\g \Dot \b} \Phi^{\rm L} 
\,) |_{pion}{~~~~~~~~~~}  \cr
 &=~ \frac 14 sin^2 (2 \g_{\rm S}) \int d^4 x\,  {\cal J}_{\rm I \, 
J \, K \, L} (\Pi) \, {\rm P}^{\un a \, \un b \, \un c \, \un d} \,  \Big[\,
( \pa_{\un a} \Pi^{\rm I} \, ) \, ( \pa_{\un b} \Pi^{\rm J} \, )
\, (\pa_{\un c} \Pi^{\rm K} \,) \, ( \pa_{\un d} \Pi^{\rm L} \,)  \, 
\Big] ~~~.   }\eqno(6.8) $$

The remaining dynamical fields in (2.8), the pionino fields, can be 
described by an SU(3) octet of Dirac spinors ($\ell_{(\a)}{}^{\rm I} 
\equiv (\psi^{\rm I}_{\a} , \, {\Bar \z}{}^{\rm I}_{\ad}$)) or using 
a 4-component spinor notation 
$$\ell^{\rm I}(x)  ~\equiv ~ 
\left(\begin{array}{c}
 \psi^{\rm I}_{\a}\\
~\\
{\Bar \z}{}^{\rm I}_{\ad} \\
\end{array}\right) ~~~~~~
\g^5 =\left(\begin{array}{cc}
~ {\rm I}_2 & ~~ 0 \\
{}~&~\\
~0 & ~~ - {\rm I}_2 \\
\end{array}\right)  {~~~~,~~~~} 
\begin{array}{c}
 \psi^{\rm I}_{\a} ~=~ \frac 12 (\,  {\rm I} + \g^5 \,) \ell \\
~\\
{\Bar \z}{}^{\rm I}_{\ad} ~=~  \frac 12 (\,  {\rm I} - \g^5 \,) \ell\\
\end{array}
~~~.
\eqno(6.9) $$
Using the definitions in (6.6, 6.7, 6.9), the action in (6.3) may be 
rewritten in terms of the Dirac spinor $\ell^{\rm I}$ and the bosons 
$(\, \Pi^{\rm I}, \, \Theta^{\rm I} ,\, {G}^{\rm I}_+ ,\,
{G}^{\rm I}_- \, )$.

\section{Toward the Gauged Auxiliary-free 4D, N = 1 
\newline  Supersymmetric QCD Effective Action}

~~~~The problem of deriving the complete gauged version of the
CNM-WZNW term will be addressed in a future work. In this section,
however, we wish to establish some fundamental structures that will
likely be required for its complete construction.  The topic of
gauging isometries in 4D, N = 1 supersymmetric non-linear $\s$-models
began sometime ago \cite{BagWit}. However, the superfield methods
needed to tackle the present problem were established by Hull, Karlhede, 
Lindstrom and Ro\v cek \cite{ISO}. It is not our purpose here to
review their formalism, however, we do note that even the gauging of 
the terms {\underline {only}} involving chiral superfields in the
K\" ahler potential of (2.4), is non-trivial. Before writing their
result, we note that equation (4.3) can also be written as
$$\d_{\xi} \Phi^{\rm I} ~=~ \a^{(A)} [ \, \xi^{\rm K}_{(A)} 
( \partder{~~~}{\Phi^{\rm K}} ) \, , \, \Phi^{\rm I} \, ]
~\equiv~ - i L_{\a}  \Phi^{\rm I} ~~~.  \eqno(7.1) $$
Now we simply note their result for gauging the pure chiral terms reads,
$$
\int d^4 x  d^2 \q \, d^2 {\Bar \q} ~ \Big[ ~ K(\Phi, {\Bar \Phi})
 ~+~ \Big(\Big( \, {{~ e^{L_V } ~-~ 1 ~} \over {L_V} }\Big)  
X_{(A)} V^{(A)} \Big)  ~\Big]
~~~, \eqno(7.2) $$
where we have used the following definitions,
$$ K(\Phi, {\Bar \Phi}) ~\equiv~ \frac 12 \, [~ {\Bar \Phi}^{\rm I} 
{\rm H}_{\rm I}({\Phi}) ~+~ {\rm {h.\, c.}}~] ~~~,~~~
\xi^{\rm I}_{(A)} ( \partder{~~}{\Phi^{\rm I}} ) K ~=~ - i X_{(A)}
~+~ \eta_{(A)}  ~~~, \eqno(7.3) $$
and the second equation above acts as the definition of the
real superfield function $X_{(A)}(\Phi, {\Bar \Phi})$ required
to write the gauging (note that in (4.7) we
make the replacement $\eta \to \eta_{(A)}$) of the isometry.
One point of interest can be seen in that this action is an infinite 
powers series in the vector fields $V^{(A)}$. This is in accord
with a theorem stated in \cite{ACG}. Namely whenever a global
``symmetry'' exists that only leaves an action invariant up to 
surface terms, the coupling to gauge field must be other
than through simple minimal covariantization. (This is the same
reason that minimal coupling does not correctly gauge the ordinary
WZNW term.) For the purely chiral superfield terms of (2.4), equation
(4.7) tells us that the isometries generate surface terms proportional
to $\eta_{(A)} + {\Bar \eta}_{(A)}$. There is something very intriguing 
about (7.2-7.3). Let us assume that the superfields $\eta$ are identically 
zero, then (7.2) can be re-written as
$$
\int d^4 x  d^2 \q \, d^2 {\Bar \q} ~  K(e^{L_V } \Phi, {\Bar \Phi})
~~~, \eqno(7.4) $$
noting that $L_V$ is defined by (7.1) with the replacement $\a \to V$.
This result looks exactly like what one would obtain in a Yang-Mills
theory! We may think of the operator superfield
$$
\exp[ V^{(A)} \xi^{\rm K}_{(A)}  (\partder{~~~}{\Phi^{\rm K}} ) ]
~\equiv ~ \exp[ V^{(A)} \xi^{\rm K}_{(A)} \pa_{\rm K}  ]
~~~, \eqno(7.5) $$
as we would any similar operator in a 4D, N = 1 supersymmetric
Yang-Mills theory. This is also supported by noting that the
algebra of $\xi^{\rm K}_{(A)}  \pa_{\rm K} $
must be closed thus this acts just as the matrix generators of
some compact Lie algebra.  With this recognition, we see that the
transformation laws in (4.3) and (4.4) may be derived as
the infinitesimal versions of the equations
$$
(\Phi^{\rm I}\,)' ~=~ \exp[ \a^{(A)} \xi^{\rm K}_{(A)}  
\pa_{\rm K}  ] \, \Phi^{\rm I} 
~~~, ~~~ $$
$$(\S^{\rm I}\,\pa_{\rm I} )' ~=~ \exp[ - \a^{(A)} \xi^{\rm K}_{(A)} 
\pa_{\rm K} ] \, (\S^{\rm I} \,\pa_{\rm I}\, ) \exp[  \a^{(A)} \xi^{\rm K}_{(A)}  
\pa_{\rm K}  ] \, 
~~~. \eqno(7.6) $$
This last observation suggests that we re-examine the issue of gauging
isometries in superspace.  We are further motivated to look at this
question particularly in light of the existence of K\" ahler
manifold covariant derivatives that were shown to exist sometime
ago \cite{kol}.

Given the operator in (7.5) it is natural to define a group of
gauge transformations acting on the superfield $V^{(\rm I)}$
via the equation
$$( \exp[ V^{(A)} \xi^{\rm K}_{(A)}\pa_{\rm K} ] \, )' 
~=~ \exp[ - i {\Bar \L}^{(A)} \xi^{\rm K}_{(A)} \pa_{\rm K} ] \, 
 \exp[ V^{(A)} \xi^{\rm K}_{(A)}  \pa_{\rm K}  ] 
\exp[ i {\L}^{(A)} \xi^{\rm K}_{(A)} \pa_{\rm K} ]
~~~. \eqno(7.7) $$
Thus there appears to be no impediment that prevents us from 
using the operator in (7.5) as the starting point in the 
construction of a K\" ahler manifold covariant derivative
that is modeled exactly like the usual 4D, N = 1 supersymmetric
Yang-Mills covariant derivative. (This construction differs
somewhat from that used by Koller.) 

Once we notice these points, the covariantization of the nonminimal
multiplet terms becomes trivial.  We use the same prescription as for
the chiral multiplets. Alternately, once we possess K\" ahler
manifold covariant deivatives, we can also accomplish the gauging
of isometries simply by replacing the derivatives in (2.1) by their
appropriate K\" ahler manifold covariant derivatives.  This superspace 
minimal coupling procedure can be used on all of the higher derivative 
terms too...with the exception of the CNM-WZNW term to which we now 
turn.

The component level ungauged WZNW term has the familiar form,
$$ \eqalign{ {~~~~}
{\cal S}_{WZNW} &=~ C_0
\int d^4 x \, \int_0^1 d y ~ {\rm {Tr}} \Big[ \, ( {\Hat U}^{-1} 
\pa_y {\Hat U} \,) ~ {\Hat {\cal W}}_4 \, \Big]   ~~~~, {~~~~~~~~~
~~~~~~~~~~~~~~~}\cr
{\Hat {\cal W}}_4 &=~ \e^{{\un a}{\un b}{\un c}{\un d}} \, (\pa_{\un a} 
{\Hat U} ^{-1} \,) \, (\pa_{\un b} {\Hat U}  
\,) \, (\pa_{\un c} {\Hat U}^{-1} \,) \,  (\pa_{\un d} {\Hat U} 
\,) \, \equiv \, (d {\Hat U} ^{-1} \,) \, (d {\Hat U}  
\,) \, (d {\Hat U}^{-1} \,) \,  (d {\Hat U} \,)
~~~~, \cr
C_0 &=~  - i N_C \, [ \, 2 {~}_{\Dot {~}} 5! \, ]^{-1} ~~~~, }   
\eqno(7.8) $$
expressed in term of the $y$-extended SU(3) group element $\Hat U = \exp 
[\, i y f_{\pi}^{-1} \Pi \,]$. For the rest of the gauged WZNW action we 
simply note that $U \equiv {\Hat U}(y = 1)$.

Using the results of ref. \cite{D1} as our starting point, we find that the 
form of the gauged WZNW action that insures the conservation of the vector 
current can be expressed as
$$ \eqalign{
S_{WZNW}^{gauged} (U, \, A_L, \, A_R ) &=~ C_0 \Big[ S_{WZNW} (U) ~+~ 
S_{WZNW}^{MC^3} (U, \, A_L, \, A_R ) \cr
&{\,~~~~~~~~}+~ S_{WZNW}^{MC^2} (U, \, A_L, \, A_R ) ~+~ S_{WZNW}^{MC^1} 
(U, \, A_L, \, A_R )  \cr
&{~\,~~~~~~}~+~ S_{WZNW}^{MC^0} (U, \, A_L, \, A_R ) ~+~ S_{WZNW}^{AF} 
(U, \, A_L, \, A_R ) \cr
&{~~~~\,~~~~}+~ S_{WZNW}^{A^2 F} (U, \, A_L, \, A_R ) ~\Big]  
~~~. }   \eqno(7.9) $$
The explicit form of $S_{WZNW}$ has already been given, so below we 
give the remaining terms
$$ \eqalign{ 
S_{WZNW}^{MC^3} (U, \, A_L, \, A_R ) ~=~ - &\int d^4 x Tr \Big[ \, 
(d U) \, (d U^{-1} ) \, (d U ) \, U^{-1} \, A_L \, \Big]  \cr
~+~ &\int d^4 x Tr \Big[ \, (d U^{-1} ) \, (d U ) \, (d U^{-1} ) 
\, U \, A_R \, \Big]  ~~~,
}    \eqno(7.10) $$

$$ \eqalign{ 
S_{WZNW}^{MC^2} (U, \, A_L, \, A_R ) ~=~  - \frac 12 &\int d^4 x 
Tr  \Big[ \, (d U) \, U^{-1} \, A_L  \, (d U) \,  U^{-1}  \, (d U ) \, 
A_L \, \Big]  \cr
+ \frac 12 &\int d^4 x  Tr  \Big[ \, U^{-1} \, (d U) \, A_R \, \, 
U^{-1}  \, (d U) \, A_R \,  \Big]  \cr
+  &\int d^4 x  Tr  \Big[ \, (d U) \,  (d U^{-1} ) \,   U^{-1}  \,
A_L \, U \,  A_R \,  ~  \Big] \cr
-  &\int d^4 x  Tr  \Big[ \, (d U) \,  (d U^{-1} ) \,A_R \, U^{-1}  
\, A_L \, U  ~  \Big]  ~~~,  
}    \eqno(7.11) $$

$$  \eqalign{ 
S_{WZNW}^{MC^1} (U, \, A_L, \, A_R ) ~=~ -  &\int d^4 x 
Tr  \Big[ \,  (d U ) \, (\,   U^{-1}  \, A_L \, A_L \, A_L ~+~ 
 A_R \, A_R \, A_R \,  U^{-1}  \, )~  \Big]     \cr
+  &\int d^4 x Tr  \Big[ \,  (d U ) \, U^{-1}  \, A_L \, U \, 
A_R \,  U^{-1}  \, A_L   ~  \Big] \cr
+  &\int d^4 x Tr  \Big[ \,  (d U ) A_R \, U^{-1} \, A_L \, U \, 
A_R \, U^{-1}  ~  \Big]    \cr
+  &\int d^4 x  Tr  \Big[ \,  (d U^{-1} ) \,  A_L \, A_L \, U \, 
A_R  ~  \Big] \cr
-  &\int d^4 x  Tr  \Big[ \, (d U ) \,  A_R \, A_R \, U^{-1} \, 
A_L  ~ \Big]  ~~~, }   \eqno(7.12)  $$

$$  \eqalign{ 
S_{WZNW}^{MC^0} (U, \, A_L, \, A_R ) ~=~ -  &\int d^4 x  Tr  \Big[ 
\, U^{-1} \, A_L \, A_L \, A_L \, U \,  A_R ~-~ U \,  A_R \, A_R 
\, A_R \, U^{-1} \,  A_L ~ \Big]   \cr 
-  &\int d^4 x  Tr  \Big[ \, U^{-1} \, A_L \, U \, A_R \, U^{-1} \, 
A_L \, U \, A_R ~  \Big] ~~~, 
}   \eqno(7.13) $$

$$  \eqalign{ 
S_{WZNW}^{AF} (U, \, A_L, \, A_R ) ~=~  +  &\int d^4 x  Tr  \Big[ \, 
(d U) \, U^{-1} \, ( \, A_L  \, F_L ~+~ F_L  \, A_L \, ) ~  \Big] \cr 
-  &\int d^4 x  Tr  \Big[ \, (d U^{-1}) \, U \,  ( \, A_R  \, F_R ~+~ 
F_R  \, A_R \, ) ~ \Big]  \cr
 +  &\int d^4 x  Tr  \Big[ \, (d U) \, F_R \, U^{-1} A_L ~-~ (d 
U^{-1} ) \, F_L \, U^{-1} A_R ~  \Big]
~~~, }   \eqno(7.14) $$

$$  \eqalign{ 
S_{WZNW}^{A^2 F} (U, \, A_L, \, A_R ) ~=~   +  &\int d^4 x  Tr  \Big[ 
\, F_L \, ( \, A_L \, U \, A_R \, U^{-1} ~-~ U \, A_R \,  U^{-1} \, 
A_L \, ) ~  \Big]  \cr  
+  &\int d^4 x  Tr  \Big[ \, F_R \, ( \,  U^{-1} \, A_L \, U \, A_R  
~-~  A_R \,  U^{-1} \, A_L \,  U \, )~  \Big]   ~~~.}   \eqno(7.15) $$

This particular decomposition is motivated only by the fact that it
makes an especially simple starting point in an effort to embed the
entirety of $S_{WZNW}^{gauged}$ within a 4D, N = 1 supersymmetric action.  
For example, the terms $S_{WZNW}^{AF}$ and $S_{WZNW}^{A^2 F}$ must
depend linearly on the Yang-Mills field strength supertensor 
$W_{\a}^{(A)}$ in a supersymmetric extension of these terms.

An important point to realize regarding (6.4) is that we should
have in mind the similarity relations between superfield pull-back
quantities and component field pull-back quantities. For example,
$$
({\Bar D}{}^{( \ad} \S^{\rm I} \, ) ({\Bar D}{}^{\Dot \b )} \S^{\rm J} \, 
) (\pa^{\g}{}_{\ad} \Phi^{\rm K} \,) ( \pa_{\g \Dot \b} \Phi^{\rm L} \,)
~\sim~ \e^{\un a \, \un b \, \un c \, \un d} (\pa_{\un a} \phi^i \, )
 (\pa_{\un b} \phi^j \, )  (\pa_{\un c} \phi^k \, )  (\pa_{\un d} 
\phi^l \, ) ~~~.
\eqno(7.16) $$
In order to write the gauged version of the WZNW action, we
require lower order pull-back quantities. For these we require
ansatz\" e and we propose these as given in the table below.
\begin{center}
\renewcommand\arraystretch{1.2}
\begin{tabular}{|c|c| }\hline
${\rm Comp. Field.~Pull-back }$ & ${\rm Superfield.~Pull-back}$
\\ \hline \hline
$ ~~ \e^{\un a \, \un b \, \un c \, \un d} (\pa_{\un b} 
\phi^j \, )  (\pa_{\un c} \phi^k \, )  (\pa_{\un d} 
\phi^l \, ) ~~$ &  $ ~~({\Bar D}{}^{\ad} \S{}^{\rm J} \, ) \, 
(\pa_{\un a} \Phi^{\rm K} \,) ({D}^{\a} {\Bar \S}{}^{\rm L} \, )  
~~$ \\ \hline
$  \e^{\un a \, \un b \, \un c \, \un d} (\pa_{\un c} \phi^k \, )  
(\pa_{\un d} \phi^l \, ) $ &  $ (\pa_{\un a} {\Bar \Phi}{}^{\rm K} \,) 
({D}^{\a} {\Bar \S}{}^{\rm L} \, ) $ \\ \hline
$  (\pa_{\un d} \phi^l \, ) $ &  $ ({D}_{\a} 
{\Bar \S}{}^{\rm L} \, ) $ \\ \hline
\end{tabular}
\end{center}
\centerline{{\bf Table IV}}
This table is not meant to be exhaustive. However, these particular
choices have some very interesting properties. 

Motivated by these choices and the discussion of super $p$-form
geometry in chapter five, we present the following ansatz\" e
for the 4D, N = 1 superfield generalizations of the terms in
equations (7.10-7.15),
$$ \eqalign{
S_{CNM-WZNW}^{MC^3} ~=  ~\int d^4 x \, d^2 \q \, d^2 
{\Bar \q} ~ &\Big[ ~ {\rm g}_{\rm {(I) \, J \, K \, L}} (\Phi , \,
{\Bar \Phi}, \,V) ~ V^{(\rm I)} \, ({\Bar D}^{\ad} \S^{\rm J} \, ) \, 
(\pa_{\un a} \Phi^{\rm K} \,) ({D}^{\a} {\Bar \S}{}^{\rm L} \, ) \cr 
&{~~~}+{\rm {h.\, c.}} ~\Big]  ~~~,
{~~~~~~~~~~~~~~~~~~~~~~~~~~~~~~~~~~~~~~~~~} (7.17) \cr
S_{CNM-WZNW}^{MC^2} ~=~  \int d^4 x \, d^2 \q \, d^2 {\Bar \q} 
~ &\Big[ ~{\rm g}_{\rm {(I) \, (J) \, K \, L}} (\Phi , \,
{\Bar \Phi},\,V) ~ V^{(\rm I)} \, \G^{\ad} {}^{\, (\rm J)} \,
(\pa_{\un a} {\Bar \Phi}{}^{\rm K} \,) ({D}^{\a} {\Bar \S}{}^{\rm L} \, )\cr
&{~~~}+{\rm {h.\, c.}} ~\Big]  ~~~,
{~~~~~~~~~~~~~~~~~~~~~~~~~~~~~~~~~~~~~~~~~} (7.18) \cr
S_{CNM-WZNW}^{MC^1} ~=~  \int d^4 x \, d^2 \q \, d^2 
{\Bar \q} ~ &\Big[ ~{\rm g}_{\rm {(I) \, (J) \, (K) \, L}} (\Phi , \,
{\Bar \Phi},\,V) ~ V^{(\rm I)} \, \G^{\ad} {}^{\, (\rm J)} \,
\G_{\un a} {}^{(\rm K)}  ({D}^{\a} {\Bar \S}{}^{\rm L} \, ) \cr
&{~~~}+{\rm {h.\, c.}} ~\Big]  ~~~,
{~~~~~~~~~~~~~~~~~~~~~~~~~~~~~~~~~~~~~~~~~} (7.19) \cr
{\cal S}_{CNM-WZNW}^{MC^0} ~=~  \int d^4 x \, d^2 \q \, d^2 
{\Bar \q} ~  &\Big[ ~{\rm g}_{\rm {(I) \, (J) \, (K) \, (L)}} (\Phi , \,
{\Bar \Phi},\,V) ~ V^{(\rm I)}\,  \G^{\a \, (\rm J)} \G_{\un a}{}^{(\rm K)} 
\G^{\ad \, (\rm L)} ~~~,{~~}{~~}{~~}\,\, \cr 
&{~~~}+{\rm {h.\, c.}} ~\Big]  ~~~,
{~~~~~~~~~~~~~~~~~~~~~~~~~~~~~~~~~~~~~~~~~} (7.20) \cr
S_{CNM-WZNW}^{AF}  ~=~ \int d^4 x \, d^2 \q \, d^2 {\Bar \q} ~ 
 &\Big[ ~{\rm g}_{\rm {(I) \, (J) \, K}} (\Phi , \, {\Bar \Phi},\,V) ~
V^{(\rm I)} \, W^{\a} \, {}^{(\rm J)} ({D}_{\a} {\Bar \S}{}^{\rm K} \, ) \cr 
&{~~~}+{\rm {h.\, c.}} ~\Big]  ~~~,
{~~~~~~~~~~~~~~~~~~~~~~~~~~~~~~~~~~~~~~~~~} (7.21) \cr
}$$
$$  \eqalign{
S_{CNM-WZNW}^{A^2F}  ~=~ \int d^4 x \, d^2 \q \, d^2 {\Bar \q} ~
 &\Big[ ~{\rm g}_{\rm {(I) \, (J) \, (K)}} (\Phi , \, {\Bar \Phi},\,V) ~
V^{(\rm I)} \, W^{\a} \, {}^{(\rm J)} \G_{\a} {}^{(\rm K)} \cr 
&{~~~}+{\rm {h.\, c.}} ~\Big]  ~~~,
{~~~~~~~~~~~~~~~~~~~~~~~~~~~~~~~~~~~~~~~~~} (7.22) } $$
where the non-polynomial functions ${\rm g} (\Phi , \, 
{\Bar \Phi},\,V)$ are yet to be determined.   Let us end this
chapter on a cautionary note. We have been guided by the fact that 
the actions above represents a ``minimal choice'' which has the 
property of producing the minimal number of component level terms.
In fact, it seems that the terms above maintain the auxiliary-freedom
for the gauged CNM-WZNW term (at least in the WZ gauge). There are 
ambiguities in making these ansatz\" e.  Any of the factors 
involving the chiral and nonminimal multiplets can be replaced by 
terms where a different scalar multiplet is used (e.g. in (7.17) 
$\S \to {\Bar \Phi}$ etc.). However, such replacements are not
guaranteed to maintain auxiliary-freedom.  This can be seen
most vividly in (7.17). Replacing both nonminimal multiplets
by chiral multiplets leads to an action that is isomorphic
to the NR-WZNW term!

\section{Contemplations on Holomorphy and Super- \newline $\,$symmetric
Yang-Mills Theory}

~~~~In the past \cite{D}, we voiced concern over the singular direction
of investigations of supersymmetric phenomenologically interesting models. 
In particular, the universal use of models wherein matter was {\underline 
{solely}} represented by chiral Wess-Zumino superfields seemed to us a 
very imprudent and incomplete course to pursue.  Although our comments
may have been viewed as vague misgivings, some implications of this
present work permit us to give a much sharper and clearer discussion 
of our concerns. These concerns will be articulated in this chapter.

Holomorphy, as we have interpreted it in this work, is a concept that
applies to the complete effective action in (2.8). The isometry group 
of the K\" ahler manifold suggested a new and intrinsic definition of 
space-time ``vector-like''  effective actions where spinors of both 
space-time chirality are found to be elements of ${}^* T_p({\cal M})$.  
Presumably in a fundamental supersymmetric version of QCD, there is no 
such K\" ahler geometry, so at first there would seem to be no way that 
the new definition of a ``vector-like'' theory might apply.  This is not 
quite true. In a fundamental theory, the role of the isometry group of 
the K\" ahler manifold is taken over by the actual supersymmetric 
Yang-Mills gauge group.  If we write the transformation laws of 4D, 
$K$-gauge N = 1 Yang-Mills supercovariant derivative $\nabla_{\un A} 
\equiv (\nabla_{\a}, \, {\Bar \nabla}_{\ad}, \, \nabla_{\un a})$ in 
the usual $K$-gauge formulation,
$$
(\nabla_{\a}, \, {\Bar \nabla}_{\ad}, \, \nabla_{\un a})' ~=~
e^{i K} \, (\nabla_{\a}, \, {\Bar \nabla}_{\ad}, \, \nabla_{\un a})\,
e^{- i K} ~~~,
\eqno(8.1) $$
with the usual hermitian parameter superfields $K$, it would seem
impossible for the concept of a ``holomorphic'' Yang-Mills gauge
group (analogous to the K\" ahler manifold isometry group) to arise.
Covariantly chiral superfields $\Phi$ certainly transform as
$$
{\Phi}' ~=~ e^{i K} \,   {\Phi}  ~~~,~~~  {\Bar 
\nabla}_{\ad} {\Phi} ~=~ 0  ~~~.   
\eqno(8.2) 
$$
However, the ``true'' gauge group of supersymmetric Yang-Mills theory 
is the $\L$-gauge group.  With respect to the $\L$-gauge group we 
define a Yang-Mills covariant $\nabla_{\un A} \equiv (\nabla_{\a}, 
\, {\Bar D}_{\ad}, \, \nabla_{\un a})$ and 
chiral superfields $\Phi$ transform as
$$
{\Phi}' ~=~ e^{i \L} \,   {\Phi} ~~~,~~~  {\Bar 
D}_{\ad} {\Phi} ~=~ 0  ~~~.   
\eqno(8.3) 
$$

With respect to supersymmetric Yang-Mills theory in superspace,
there are {\underline {three}} Yang-Mills type fiber bundles over 
the supermanifold, one associated with each $K$, $\L$ and ${\Bar \L}$.
For a given Yang-Mills group denoted by $G$, let us denote the
three fiber bundles by ${\cal F}_K (G)$, ${\cal F}_{\L} (G)$
and ${\Bar {\cal F}}_{\L} (G)$ respectively.  It is abundantly clear
that ${\cal F}_{\L} (G)$ is holomorphic and ${\Bar {\cal F}}_{\L} (G)$
is anti-holomorphic.

It has been well known that the $K$-gauge is an artifact, so we should 
really focus our attention on ${\cal F}_{\L} (G)$ and ${\Bar {\cal F}}_{\L} 
(G)$.  Since for the spinors in the chiral and conjugate chiral multiplet
their respective transformations yield
$$ \eqalign{
(\psi_{\a})' &=~ (\nabla_{\a} \Phi)' ~=~ e^{i \hat \L} (\nabla_{\a} \Phi)
~=~ e^{i \hat \L} \psi_{\a} ~~~, \cr
({\Bar \psi}_{\ad})' &=~ ({\Bar \nabla}_{\ad} {\Bar \Phi})' ~=~ e^{-i \hat 
{\Bar \L}} ({\Bar \nabla}_{\ad} {\Bar \Phi})
~=~ e^{- i \hat {\Bar \L}} {\Bar \psi}_{\ad} ~~~, }
\eqno(8.4) $$
(where $\hat \L$ and $\hat {\Bar \L}$ denote the $\q = 0$ part of the
superfield) we find
$$
\psi_{\a} ~ \in ~ {\cal F}_{\L} (G) ~~~,~~~ {\Bar \psi}_{\ad} \in 
{\Bar {\cal F}}_{\L} (G) ~~~. \eqno(8.5) $$
Thus, there exists the same type of correlation between the space-time
chirality and a holomorphic or anti-holomorphic structure.  The only
way to break this correlation is to introduce non-minimal multiplets.

Covariant nonminimal superfields $\S$ transform as
$$
{\S}' ~=~ e^{i K} \,   {\S}  ~~~,~~~  {\Bar 
\nabla}^2 {\S} ~=~ 0  ~~~,   
\eqno(8.6) 
$$
or with respect to the $\L$-gauge group as
$$
{\S}' ~=~ e^{i \L} \,  {\S} ~~~,~~~  {\Bar 
D}^2 {\S} ~=~ 0  ~~~.   
\eqno(8.7) 
$$
The physical spinors in the nonminmal multiplets transform as
$$ \eqalign{
({\Bar \z}_{\ad})' &=~ ({\Bar \nabla}_{\ad} {\Bar \S})' ~=~ e^{i \hat 
{\L}} ({\Bar \nabla}_{\ad} {\Bar \S}) ~=~ e^{i \hat \L} {\Bar \z}_{
\ad} ~~~,  \cr
(\z_{\a})' &=~ (\nabla_{\a} {\Bar \S})' ~=~ e^{-i \hat {\Bar \L}} 
(\nabla_{\a} {\Bar \S})
~=~ e^{-i \hat {\Bar \L}} \z_{\a} ~~~, \cr
 }
\eqno(8.8) $$
so that we immediately see 
$$
{\Bar \z}_{\ad} ~ \in ~ {\cal F}_{\L} (G) ~~~,~~~ {\z}_{\a} \in 
{\Bar {\cal F}}_{\L} (G) ~~~. \eqno(8.9) $$

If these notions are correct, the reader may then wonder, ``In what
sense is the standard construction, used throughout the literature,
vector-like?''  From our view the simplest answer to this is that
the standard construction is vector-like with respect to ${\cal 
F}_{K} (G)$.  However, the inability to write an auxiliary-free
WZNW term with respect to this structure raises our concern that
there may be subtle difficulties in such an approach.

Another way to formulate these issues is based on the following
argument. Let $\Psi$ denote a Dirac spinor field.  Our discussion
indicates that there are two ways in which to construct such an
object from 4D, N = 1 superfields. One definition, we
call the C-definition (chiral superfield), is given by
$$ \Psi_C (x)~\equiv~ \left(\begin{array}{c}
 D_{\a} \Phi_+ | \\
~\\
{\Bar D}_{\ad} {\Bar \Phi}_- | \\
\end{array}\right)  ~~~, 
\eqno(8.10) $$
with $\Phi_+ \neq \Phi_-$. If $\Phi_+ = \Phi_-$, then ($\Psi_C)^*
= \s^1 \Psi_C$ which implies that $\Psi_C$ defines a Majorana fermion.  
The undotted components of a Dirac spinor reside in $\Phi_+$ while
the charge conjugates of the dotted components reside in $\Phi_-$.
The other definition, referred to as the CNM-definition (chiral-nonminimal
superfield), is given by
$$ \Psi_{CNM} (x) ~\equiv~ \left(\begin{array}{c}
 D_{\a} \Phi | \\
~\\
{\Bar D}_{\ad} {\S} | \\
\end{array}\right) ~~~.
\eqno(8.11) $$
In (8.10) and (8.11) it is to be understood that the $\q \to 0$ limit 
must be taken on the superfields.  The coupling of these Dirac fields to 
a U(1) gauge superfield $V$ (we consider U(1) for the sake of simplicity) 
is given by the respective Lagrangians
$$ {\cal L}_C ~=~ {\Bar \Phi}_+ e^V  \Phi_+ ~+~ {\Phi}_- e^{- V}  
{\Bar \Phi}_- ~~~, $$
$$ {\cal L}_{CNM} ~=~ {\Bar \Phi} e^V  \Phi ~-~
{\Bar \S} e^V  \S ~~~.  \eqno(8.12) $$
Now an interesting point is to compare how the U(1) gauge symmetry
is realized on the two different definitions for a Dirac field.
$$(\Psi_C )' ~=~ \exp \Big[ \, i \frac 12 ({\hat \L} + {\hat {\Bar \L}}) 
~+~ i \frac 12 \s^3 ({\hat \L} - {\hat {\Bar \L}} ) \, \Big] \, \Psi_C 
~~~,~~~ $$
$$
(\Psi_{CNM} )' ~=~ \exp \Big[ \, i {\hat \L} \, \Big] \Psi_{CNM} ~~~.
\eqno(8.13) $$
This equation illustrates the point that for Dirac spinors 
$\Psi_{CNM}$ or $\Psi_C$ contained in a 4D, N = 1 supersymmetric theory 
(outside of the Wess-Zumino gauge) the gauge group is not $G$ as in 
component theories, but is either $G_c$ (the complexification of $G$) 
or $G_V \otimes G_A$.  In a Wess-Zumino gauge ${\hat \L} (x) = {\hat 
{\Bar \L}} (x)$ and these two definitions for the U(1) transformation 
law of a Dirac particle coincide. It is completely clear that the group 
of gauge transformations only for $\Psi_{CNM}$ form a holomorphic U(1) 
group. For $\Psi_C$, the term dependent on $i({\hat \L} - {\hat {\Bar 
\L}} )$ is a $\g^5$-rotation (${\cal R}$-symmetry transformation)!  This 
raises a real possibility of having an auxiliary-field anomaly\footnote{This 
is more often referred to as the ``Konishi anomaly'' \cite{KKon} although the
first discussion of this \newline ${~~~~~}$ phenomenon in the physics 
literature was in ref. \cite{GGS}.} \cite{GGS}.

\section{ Conclusion}

~~~~We believe our recent realizations have a significance for a disagreement
that occurred some years ago.  At that time, we made the first explicit
suggestion \cite{AAA,CCC} as to how the string corrections associated with 
the presence of the Lorentz Chern-Simons form modify the geometry of 10D, 
N = 1 superspace supergravity.   Later a different suggestion \cite{BPT} 
BPT-FFP appeared in the literature and there ensued a vigorous disagreement over 
which approach was ``the correct one.''  Although most theorists conversant 
in this matter concluded that the approach of \cite{BPT} was correct, we
remain absolutely convinced that our suggestion of \cite{AAA} is indeed
the correct one.  We would like to note some extremely interesting analogies
regarding the disagreements of the works of \cite{B2} versus \cite{SG} and 
those of \cite{AAA,CCC} versus \cite{BPT}.

Foremost, note that our equation of (2.8) is ``spectrum stable.''  By this
we mean that independent of the order to which we expand in $\g'$,
the spectrum of dynamical fields in the action remains unchanged.  This
condition is violated if we replace the leading term in ${\cal S}_{{\rm H}.\,
{\rm D}.}$ by the BNS and NR actions. To lowest order in $\g'$ only
the $A$ and $B$ fields propagate and at higher order the $F$ fields propagate.

Let us now contrast this with the results of \cite{AAA,CCC} versus \cite{BPT}.
In our previous work on the 10D, N  = 1 superspace geometry associated with 
the low-energy effective action of the heterotic strings \cite{AAA}, both
at zero order and first order in $\g'$, the spectrum is unchanged.  By way of
comparison, in the superspace geometry associated with the low-energy 
effective action of the heterotic strings according to \cite{BPT}, the 
spectrum at zero order in $\g'$ is different from that at first order in 
$\g'$. In particular, the approach of \cite{BPT} requires for its 
mathematical consistency a propagating spin-connection.  However, this 
spin-connection only propagates when the first order $\g'$-terms are 
included. At zero order in $\g'$, the BPT-FFP spin-connection has an
algebraic equation of motion.  This is just like the explicit example we 
showed by studying the behavior of the $F$-field in either the BNS or 
NR higher derivative actions.   

Thus by analogy, we assert that the work of \cite{BPT} is the 10D, N  = 1 
superspace geometry associated with a higher dimensional generalization of 
an action in the same class as the BNS or NR actions.  This would mean 
that the mathematical correctness of the BTP-FFP approach (which we previously 
doubted) is {\underline {no}} guarantee of uniqueness!  Also by analogy, we 
assert that our work in reference \cite{AAA} is the higher dimensional 
generalization of the class of actions we have discussed here as well as 
in \cite{SG}.  We are well aware of the claims by the proponents of the 
BTP-FFP approach that there is an ``obstruction'' \cite{DZ} at second 
order which precludes the higher extension of our first order results.  We 
now appeal to history.  We have always felt that the ``obstruction'' is 
really a statement about what assumptions are being made in the attempt to 
find a solution to the Chern-Simons modified 10D, N = 1 superspace geometry.  
Note this is analogous to what we have found for the auxiliary-free higher 
derivative 4D, N = 1 supersymmetric actions. If we assume that only chiral 
multiplets describe the matter fields, we are inevitably led to propagating 
$F$-fields. If we release this assumption (i.e. assign matter to both chiral 
and nonminimal multiplets) then there exists a mechanism for finding spectrum 
stable theories.  Ten years elapsed between the work of \cite{AAA,CCC} and 
\cite{SG}.  It remains to be seen what is the precise and subtle mechanism 
that would allow the existence of a spectrum stable 10D, N = 1 Lorentz 
Chern-Simons modified supergeometry that describes the low-energy effective 
action of the heterotic (and as well type-II) superstrings\footnote{We are 
hopeful that it will not take another decade to find this mechanism.}.

Perhaps the most interesting point of our suggestion of an auxiliary-free 
4D, N = 1 supersymmetric low-energy effective QCD action (and a feature
shared by the chiral-nonminimal models in references \cite{SG,D}) is that
these model provide a new way in which parity non-conservation may be 
realized.  In all models prior to the construction of these, the
mechanism for parity breaking was the inequality (either of matter fields, 
gauge fields or both) of the realization of the left Yang-Mills gauge 
group versus the right Yang-Mills gauge group on physical fields.  The 
models in our previous \cite{SG,D}) and present works show that even 
if the left Yang-Mills gauge group is equal to the right Yang-Mills 
gauge group on propagating fields, parity can still be broken in some 
circumstances by simply assigning right-handed spin-1/2 particles and 
left-handed spin-1/2 particles to different supersymmetry representations 
(i.e. chiral versus nonminimal multiplets).  Looking at the fields in 
Table I, we see that the auxiliary fields $\r_{\a}$, $p_{\un a}$ and 
${\Bar \b}_{\ad}$ transform covariantly under the left gauge group 
but have no analogs that transform covariantly under the right 
gauge group. Thus when utilized in a fundamental theory, the 
$(\Phi, \, \S)$ Poincar\' e dual pair implies a broken parity
symmetry even though parity is realized on the propagating fields
in the multiplets.  To our knowledge our present model is the first
one where the breaking of parity is {\underline {required}} by a 
theoretical reason (i.e. auxiliary freedom) of the effective
4D, N = 1 supersymmetric QCD low-energy action.

The use of our holomorphic auxiliary-free 4D, N = 1 supersymmetric low-energy
effective QCD action as a model for the real world offers a rather stark 
trade-off.  In the pursuit of the goal of achieving auxiliary-freedom 
(which is accomplished) we are forced to use a class of models in which 
the supersymmetric version of the strong interaction {\underline {breaks}} 
parity in a new way.  We approach this radical new idea with caution but 
it certainly gives us interesting new questions to explore.  Only time will 
tell if this notion, like that of supersymmetry itself, is one that is 
pleasing to Nature.

$${~~~~}$$

$${~~~~}$$

$${~~~~}$$

$${~~~~}$$

$${~~~~}$$
\noindent
{\bf {Acknowledgment; }} \newline \noindent
We wish to thank L. Rana for useful discussions. Furthermore,
we wish to acknowledge the organizers of Second International Sakharov
Conference on Physics in Moscow, Russia where some of the present
investigation was carried out and preliminary results
presented.

\newpage \noindent
{\bf {Appendix A: Chiral Superfield Maurer-Cartan Forms }}
 
We begin by interchanging the definition of left and right
Maurer-Cartan forms (from that used in \cite{SG}) motivated 
by the fact that this change simplifies later notation in the
gauged WZNW term.  Our new definition of left ($ L_m {}^i (\Pi)$) 
and right ($ R_m {}^i (\Pi)$) Maurer-Cartan forms are given by the 
equations 
$$
U^{-1} \pa_{\underline a} U ~=~ i  {f_{\pi}}^{-1} ( \, \pa_{\underline a} \Pi^m  
\,)~ R_m {}^i (\Pi) \, \l_i  ~~~~,~~~~ (\, \pa_{\underline a} U \,)  U^{-1} ~=~
i {f_{\pi}}^{-1} ( \, \pa_{\underline a} \Pi^m  \,) ~L_m {}^i 
(\Pi) \, \l_i  ~~~~.
\eqno(A.1) $$
These definitions allow $L_m {}^i (\Pi)$ and $R_m {}^i (\Pi)$ to be calculated
as power series in $\Pi^i$ from \cite{ACG}
$$\eqalign{
R_m {}^i (\Pi) &\equiv ~ (C_2)^{-1} {\rm {Tr}} \Big[\,  T^i \Big(
\frac { 1 ~-~ e^{-\D} }{\D} \Big) T_m \Big] ~~~, \cr
L_m {}^i (\Pi) &\equiv ~ (C_2)^{-1} {\rm {Tr}} \Big[\,  T^i \Big(
\frac { e^{\D} ~-~ 1 }{\D} \Big) T_m \Big] ~~~, }
\eqno(A.2) $$
where $\D T_m \equiv  i {f_{\pi}}^{-1} [ \Pi \, , \, T_m  ]$, $\D^2 T_m = \D \D 
T_m$, etc. and the constant $C_2$ is determined so that $ L_m {}^i (0) = R_m {}^i 
(0) = \d_m {}^i$.   

Next to extend these definitions to chiral superfields, we first define
group elements by 
$$ U (\Phi)  ~\equiv~ \exp \Big[ {{~{\Phi}~}\over{~ 
f_{\pi}\, cos (\g_{\rm S}) ~}} \Big] ~~~,~~~ {\Phi} ~\equiv~
{\Phi}^{\rm I} \l_{\rm I} ~~~.
\eqno(A.3) $$
Note that since the chiral superfield is complex, the quantity $U$ lies
in the complexification of the group associated with the Lie algebra
generated by $\l_{\rm I}$. Thus, $U$ is not a unitary matrix. This
definition of $U$ also has the property that when we set all of the
fields with the exception of $\Pi$ to zero, then $U$
reduces back to the usual representation involving the SU(3) pion 
multiplet.  We define our chiral superfield Maurer-Cartan forms
$ R_{\rm I} {}^{\rm K} (\Phi) $ and $ L_{\rm I} {}^{\rm K} (\Phi) $ by
$$ \eqalign{
U^{-1} D_{\a} U &=~   {[f_{\pi} \, cos (\g_{\rm S})]}^{-1} ( \, D_{\a} 
\Phi^{\rm I} \,)~ R_{\rm I} {}^{\rm K} (\Phi) \, \l_{\rm K} ~~~~,\cr
(\, D_{\a}U \,)  U^{-1} &=~  {[f_{\pi} \, cos (\g_{\rm S})]}^{-1} ( \, 
D_{\a} \Phi^{\rm I} \,) ~L_{\rm I} {}^{\rm K} (\Phi) \, \l_{\rm K}  ~~~~. }
\eqno(A.4) $$
These lead to the same expressions for $ R_{\rm I} {}^{\rm K} (\Phi) $ and 
$ L_{\rm I} {}^{\rm K} (\Phi) $ as in (A.2) except with the replacements
$\Pi \to \Phi$ and $\Delta T_{\rm I} \to {[f_{\pi} \, cos (\g_{\rm S})
]}^{-1} [ \Phi \, , \, T_{\rm I} ]$. Since the multiplication of chiral 
superfields is closed we also observe,
$$
U^{-1} {\Bar D}_{\ad} U ~=~ 0 ~~~,~~~ U {\Bar D}_{\ad} U^{-1} ~=~ 0
~~~\to ~~~{\Bar D}_{\ad}  R_{\rm I} {}^{\rm K} ~=~ 
{\Bar D}_{\ad}  L_{\rm I} {}^{\rm K} ~=~ 0~~~.
$$
$${\Bar D}_{\ad} \Big[ U^{-1} D_{\a} U \,\Big] ~=~ i
{[f_{\pi} \, cos (\g_{\rm S})]}^{-1} ( \, \pa_{\un a} 
\Phi^{\rm I} \,)~ R_{\rm I} {}^{\rm K} (\Phi) \, \l_{\rm K} ~~~~,
$$
$${\Bar D}_{\ad} \Big[ \Big(D_{\a} U \,\Big) U^{-1} \Big] ~=~ i
{[f_{\pi} \, cos (\g_{\rm S})]}^{-1} ( \, \pa_{\un a} 
\Phi^{\rm I} \,)~ L_{\rm I} {}^{\rm K} (\Phi) \, \l_{\rm K} ~~~~.
\eqno(A.5) $$

Using the real parameter $y$, we again define an extended group element 
$\Hat U$ through the relation $\Hat U = \exp [\, i y {[f_{\pi} \, cos 
(\g_{\rm S})]}^{-1} \Phi \,]$. This further implies that the Vainberg 
technique \cite{V} can be used in 4D, N = 1 superspace. If
we start with the Novikov-Witten observation \cite{CC} and write the component
action in the form,
$$ \eqalign{ {~}
{\cal S}_{WZNW} &=~ \int d^4 x \,{\cal L}_{WZNW} ~=~
- i N_C \, [ \, 2 {~}_{\Dot {~}} 5! \, ]^{-1}
\int d^4 x \, \int_0^1 d y ~ {\rm {Tr}} \Big[ \, ( {\Hat U}^{-1} \pa_y {\Hat U} 
\,) ~ {\Hat {\cal W}}_4 \, \Big]   ~~~~, \cr
{\Hat {\cal W}}_4 &=~ \e^{{\underline a}{\underline b}{\underline c}{\underline 
d}} \, (\pa_{\underline a} {\Hat U} ^{-1} \,) \, (\pa_{\underline b} {\Hat U}  
\,) \, (\pa_{\underline c} {\Hat U}^{-1} \,) \,  (\pa_{\underline d} {\Hat U} 
\,) \,   ~~~~. }  \eqno(A.6) $$
or directly using the elements of the pion octet as
$$ \eqalign{ {~~~~} 
{\cal S}_{WZNW} &=~ \int d^4 x \, \e^{{\underline a}{\underline b}{\underline c}
{\underline d}}{\cal J}_{m \, n \, r \, s} (\Pi) 
(\pa_{\underline a} \Pi^m \,) \, (\pa_{\underline b} \Pi^n \,) \,
(\pa_{\underline c} \Pi^r \,)  \, (\pa_{\underline d} \Pi^s \,) ~~~~, \cr
{\cal J}_{m \, n \, r \, s} (\Pi) &=~ - \, {N_C}  [ \, 8 {~}_{\Dot {~}} 5! 
f_{\pi}^5 \, ]^{-1} f_{a \, b}{}^k \, f_{c \, d}{}^l {\rm {Tr}} \Big[ \, 
\l_k \l_l \l_h \, \Big] \, \int_0^1 d y ~ y^4 \, \Pi^e {\Hat L}_e {}^h 
{\Hat L}_m {}^a {\Hat L}_n {}^b {\Hat L}_r {}^c {\Hat L}_s {}^d  ~~~~. }
\eqno(A.7) $$
with ${\Hat L}_m {}^i \equiv L_m {}^i (y \Pi)$ and $f_{a \, b}{}^k$ denoting
the structure constants of the group, we can simply make the replacements
$ U(\Pi) \to U(\Phi) $, $ {\Hat U}(\Pi) \to {\Hat U}(\Phi)$,  etc.~in all the 
quantities in (A.6) and (A.7). Under this circumstance all such expressions
become superfields!
$$ \eqalign{ {~~~~} 
{\cal S}_{WZNW} &=~  - i \, {N_C} \int d^4 x \, \e^{{\un a}{\un b}{\un c}
{\un d}}{\cal J}_{\rm {M \, N \, R \, S}} (\Phi) 
(\pa_{\underline a} \Phi^{\rm M} \,) \, (\pa_{\underline b} \Phi^{\rm N} \,) \,
(\pa_{\underline c} \Phi^{\rm R} \,)  \, (\pa_{\underline d} \Phi^{\rm S} \,) ~~~~, \cr
{\cal J}_{\rm {M \, N \, R \, S}}^{WZNW} (\Phi) &=~ [ \, 8 {~}_{\Dot {~}} 5! 
f_{\pi}^5 cos^5 (\g_{\rm S})\, ]^{-1} f_{\rm {A \, B \, C \, D \, E}}
\, \int_0^1 d y ~ y^4 \, \Phi^{\rm F} {\Hat L}_{\rm F} {}^{\rm E} 
{\Hat L}_{\rm M} {}^{\rm A} {\Hat L}_{\rm N} {}^{\rm B} {\Hat L}_{\rm R} 
{}^{\rm C} {\Hat L}_{\rm S} {}^{\rm D} ~~~~, \cr 
f_{\rm {A \, B \, C \, D\, E}} &\equiv~  \frac 12 f_{\rm {A \, B}}{}^{\rm K} 
\, f_{\rm {C \, D}}{}^{\rm L} {\rm {Tr}} \Big[ \, (\, \l_{\rm K} \l_{\rm L}
\, + \,  \l_{\rm L} \l_{\rm K} \,) \l_{\rm E} \, \Big] 
~~~.}
\eqno(A.8) $$

One might think that the action in (A.8) is suitable to act as the 4D, N = 1
WZNW term.  The only problem with this interpretation is that the leading
component of a superfield is not a super invariant and the component
level WZNW term is the leading component here. However, the discussion 
above does show that the holomorphic tensor ${\cal J}_{\rm {I \, J \, K \, 
L}} (\Phi)$ exists as a simple generalization of the component level
result.

We can further use this result to fix the normalization of our previous
work. Namely, the correct normalization of the CNM-WZNW action described
in \cite{SG} is
$$ \eqalign{
{\cal S}_{WZNW} ~=~ i\,  4 N_C \Big[ \frac{cos^5 (\g_{\rm S})}{sin^2 (2 
\g_{\rm S})} \Big] &\int d^4 x \, d^2 \q ~ {\cal J}_{\rm {I \, J \, K \, 
L}}^{WZNW} (\Phi) \, ({\Bar D}^{\ad} \S^{\rm I} \, ) \, ({\Bar D}^{\Dot 
\b} \S^{\rm J} \, ) \,(\pa^{\g} {}_{\ad} \Phi^{\rm K} \,) \, ( \pa_{\g 
\Dot \b} \Phi^{\rm L} \,) \cr &+~ {\rm {h. \, c.}} ~~~. }
\eqno(A.9) $$

As noted before, the action of (A.9) also contains the Skyrme term. Since
we have developed all the necessary ``technology,'' it is also simple for
us to complete the discussion of its embedding.  In (3.1) we replace every 
field by the appropriate superfield on the first line and calculate using
the chiral superfield Maurer-Cartan forms. A brief calculation reveals
that the super Skyrme term takes the form
$$ \eqalign{
{\cal S}_{WZNW}^{Skyrme} ~=~ 4 \Big[ \frac{cos^4 (\g_{\rm S})}{sin^2 (2 
\g_{\rm S})} \Big] &\int d^4 x \, d^2 \q ~ {\cal J}_{\rm {I \, J \, K \, 
L}}^{Skyrme} (\Phi) \, ({\Bar D}^{\ad} \S^{\rm I} \, ) \, ({\Bar D}^{\Dot 
\b} \S^{\rm J} \, ) \,(\pa^{\g} {}_{\ad} \Phi^{\rm K} \,) \, ( \pa_{\g 
\Dot \b} \Phi^{\rm L} \,) \cr &+~ {\rm {h. \, c.}} ~~~. }
\eqno(A.10) $$
where ${\cal J}_{\rm {I \, J \, K \,L}}^{Skyrme}$ expressed in terms
of chiral superfield Maurer-Cartan forms is given by
$${\cal J}_{\rm {I \, J \, K \, L}}^{Skyrme} (\Phi) ~=~ \Big( {1 
\over 32 e^2} \Big) [ \,
f_{\pi}^4 cos^4 (\g_{\rm S})\, ]^{-1} f_{\rm {M \, N}}{}^{\rm A}
f_{\rm {R \, S}}{}^{\rm A} \, {L}_{\rm I} {}^{\rm M} {L}_{\rm J} 
{}^{\rm N} {L}_{\rm K} {}^{\rm R} {L}_{\rm L} {}^{\rm S}  ~~~~.
\eqno(A.11) $$
Thus to include both Skyrme and WZNW terms we write (6.1) where 
${\cal J}_{\rm {I \, J \, K \,L}}$ is identified as 
$$
{\cal J}_{\rm {I \, J \, K \,L}}(\Phi) ~=~ 4 \Big[ \frac{cos^4 (\g_{\rm S}) 
\,}{sin^2 (2 \g_{\rm S})} \Big]\, \Big[ ~ {\cal J}_{\rm {I \, J \, K \, 
L}}^{Skyrme} (\Phi) ~+~  i\,  N_C \, cos (\g_{\rm S}) \, {\cal J}_{\rm {I \, 
J \, K \, L}}^{WZNW} (\Phi)~\Big] ~~~.
\eqno(A.12) $$
As expected the Skyrme and WZNW terms are seen to be the ``real'' and
``imaginary'' part of a single quantity. We put quotes around real
and imaginary since the ${\cal J}$ functions are all holomorphic.

One final item of interest is more explicit information on the holomorphic
isometry vectors. If we let SU${}_R$(3) act by right multiplication
on $U(\Phi)$ with the usual SU(3) matrices and SU${}_L$(3) act by left 
multiplication on $U(\Phi)$ with the usual SU(3) matrices, then we may
write
$$
\d U(\Phi) ~=~ i \Big[ \, \a^{({\rm I})} U(\Phi) \l_{\rm I} ~-~
 {\Tilde \a}^{({\rm I})} \l_{\rm I} \, U(\Phi)  ~  \Big] ~~~,
\eqno(A.13) $$
where $\a^{({\rm I})}$ and ${\Tilde \a}^{({\rm I})}$ are the
parameters of the transformations.  Using (A.4) these lead to the
following two equivalent expressions for the isometry vectors,
$$ \eqalign{
\xi^{\rm I} (\Phi) &\equiv ~  i [ \, f_{\pi} cos (\g_{\rm S}) \, ]
\, \Big\{ {\a}^{({\rm J})} \d_{\rm J}{}^{\rm K} ~-~ (C_2 )^{-1} 
{\Tilde \a}^{({\rm J})} \, Tr[ \, U^{-1} \l_{\rm J} U \l^{\rm K} 
\,] ~  \Big\} (R^{-1} )_{\rm K} {}^{\rm I} ~~~, \cr
&\equiv ~ - i [ \, f_{\pi} cos (\g_{\rm S}) \, ]
\, \Big\{ {\Tilde \a}^{({\rm J})}  \d_{\rm J}{}^{\rm K} ~-~ 
(C_2 )^{-1} {\a}^{({\rm J})} \,  Tr[ \, U \l_{\rm J} U^{-1} 
\l^{\rm K} \,] ~  \Big\} (L^{-1} )_{\rm K} {}^{\rm I}
~~~. } \eqno(A.14) $$
It can be seen that we also have the identities,
$$
U(\Phi) \l_{\rm I} U(- \Phi) ~=~ \Big\{ [exp (\D )] \l_{\rm I}\, \Big\}
~~~,~~~  U(- \Phi) \l_{\rm I} U(\Phi) ~=~ \Big\{ [exp (- \D )] \l_{\rm 
I}\, \Big\}~~~.
\eqno(A.15) $$
$${~~~}$$
\noindent
{\bf {Appendix B:  4D, N = 1  Superspace Formulation of
Donaldson-Nair-Schiff \newline ${~~~~~~~~~~~~~~~~~}$ Models}}

~~~It was recently proposed that the exists a class of 4D, N = 1 
supersymmetric models with some remarkable quantum properties \cite{KET}. 
In this extremely brief appendix, we wish to point out that the nonminimal 
multiplet seems ideally suited for constructing the supersymmetric extension 
of DNS models. Before we begin, let us note that the DNS model as
described in reference \cite{KET} is constructed in Atiyah-Ward space.
As noted in a previous work \cite{GKN}, all the normal machinery of
4D, N = 1 superspace can be extended into an Atiyah-Ward superspace.
Therefore we will actually give all of our arguments within the
context of 4D, N = 1 superspace. To carry out the same constructions
in Atiyah-Ward superspace merely amounts to a few minor changes
of notation.

There several elements that we need for our construction. First,
we introduce the `strung-out' version of supergravity that is
obtained as the limit of the heterotic string. This theory has been
discussed in complete detail in reference \cite{GMOV} and this allows
us to introduce supergravity supercovariant derivatives $(\nabla_{\a}, \,
{\Bar \nabla}_{\ad}, \, \nabla_{\un a})$. As explained in 
ref. \cite{GMOV}, this supergravity supercovariant derivative
gauges ${\cal R}$-symmetry with a composite connection constructed
from $G \, = \, \exp [ \varphi ]$ (the field strength of the
axion-dilaton multiplet) and its derivatives.  We note that the
super 2-form $B_{\un A \, \un B}$ can be totally constructed in 
terms of a chiral spinor prepotential \cite{PFM} that we denote
by $\varphi^{\a}$. With this information the steps to constructing
a supersymmetric DNS model are clear.

We introduce Poincar\' e dual pairs $(\Phi^I , \, \S^I )$ and replace
the $\s$-model action in (2.3) by
$$
{\cal S}_{\s}^{DNS} ~=~  \int d^4 x d^2 \q d^2 {\Bar \q} 
~ E^{-1} \, f(G) \, {\Hat \O} 
(\, \Phi, {\Bar \Phi} ;  \S, {\Bar \S} \, ) ~~~~,
\eqno(B.1) $$
where $f(G)$ is a function that must be determined by further considerations.
We next need a term that is very similar to a 2D WZNW term. Were we in a 2D 
theory, twisted chiral multiplets might be used. However, in 4D the 
alternative is to use the nonminimal multiplets.  We can construct 
this term by first introducing a holomorphic second rank anti-symmetric 
tensor in the space of the Poincar\' e dual pairs that we
denote by $b_{I \, J}(\Phi)$. This quantity is then used to write
the following holomorphic action,
$$
{\cal S}_{WZNW}^{DNS} ~=~  \int d^4 x d^2 \q {\cal E}^{-1} ~ \varphi^{\a}
\, (\nabla_{\un a} \Phi^I \, ) \, ( {\Bar \nabla}^{\ad} \S^J \, )
 ~ b_{I \, J}(\Phi) ~~~.
\eqno(B.2) $$
In this expression we have used the notation ${\cal E}^{-1}$ to denote
the usual chiral density measure. The chirality of the integrand in
this expression depends on some unusual properties of the
``heterotic 4D, N = 1 superspace supergravity derivative.''
In particular when acting on a scalar with vanishing U(1) 
${\cal R}$-charge we find 
$$
 [ {\Bar \nabla}_{\ad} \, , \nabla_{\un b} \} \Phi^I ~=~ 0 ~~~,
\eqno(B.3) $$
which is the critical condition needed to prove the chirality of the 
integrand. If we work in ordinary 4D, N = 1 superspace, we must add 
the hermitian conjugate to this last action. However, in Atiyah-Ward 
superspace everything is real and so the term above may be used 
as it stands. However, whether we use this action or its ``dotted'' 
analog has an important consequence. The duality or anti-duality 
of the spacetime 2-form gauge field that appears after reducing 
this to components is correlated with whether we use (B.2) our its 
``dotted'' analog. 

The complete 4D, N = 1 supersymmetric DNS model is just the sum of 
(B.1) and (B.2). Its construction demonstrates that without the 
existence of the nonminimal multiplet, there is {\underline {no}} 
way to describe a supersymmetric extension of the DNS model. In 
other words, the DNS model does not even have a supersymmetric 
extension if we only utilize chiral multiplets!

\newpage

\end{document}